\documentclass[
 10pt,
 twocolumn,
 superscriptaddress,
nofootinbib,
 amsmath,amssymb,
 aps,
 pra,
]{revtex4-1}

\usepackage{xcolor}
\usepackage{graphicx}% Include figure files
\usepackage{bm}% bold math
\usepackage{bbm}% math symbols
\usepackage{hyperref}% add hypertext capabilities

\usepackage{qcircuit}

\newcommand{\e}{\,\mathrm{\bf e}}

\begin{document}

\title{Estimating the gradient and higher-order derivatives on quantum hardware}

\author{Andrea Mari}
\affiliation{Xanadu, Toronto, ON, M5G 2C8, Canada}
\affiliation{Unitary Fund, CA, USA}
\author{Thomas R. Bromley}
\author{Nathan Killoran}
\affiliation{Xanadu, Toronto, ON, M5G 2C8, Canada}

%\date{\today}

\begin{abstract}
For a large class of variational quantum circuits, we show how arbitrary-order derivatives can be analytically
evaluated in terms of simple {\it parameter-shift rules}, i.e., by running the same circuit with different shifts of the parameters.
As particular cases, we obtain parameter-shift rules for the Hessian of an expectation value and for the metric tensor of a variational state, 
both of which can be efficiently used to analytically implement second-order optimization algorithms on a quantum computer.
We also consider the impact of statistical noise by studying the mean squared error of different derivative estimators. 
In the second part of this work, some of the theoretical techniques for evaluating quantum derivatives are applied to their typical use case: the implementation of quantum optimizers.
We find that the performance of different estimators and optimizers is intertwined with the values of different hyperparameters, such as a step size or a number of shots. 
Our findings are supported by several numerical and hardware experiments, including an experimental estimation of the 
Hessian of a simple variational circuit and an implementation of the Newton optimizer.
\end{abstract}

\maketitle

%\tableofcontents

%Analytic derivatives in VQE: 
%\cite{mitarai2020theory}.

%Rotosolve and 3 degrees of freedom per gate: 
%\cite{ostaszewski2019quantum}.

%Classical finite difference:
%\cite{gill2019practical, mathur2012analytical}.

%Parameter-shift rule for more general gates: 
%\cite{banchi2020measuring,crooks2019gradients}.

\section{Introduction}

Many algorithms for near-term quantum computers are based on variational
circuits \cite{peruzzo2014variational, schuld2020circuit, perdomo2018opportunities, mcclean2016theory, sim2019expressibility, killoran2019continuous}, 
i.e., sequences of quantum gates depending on classical parameters that are recursively optimized for solving specific problems. Some of the most important examples are
the variational quantum eigensolver (VQE) \cite{kandala2017hardware, peruzzo2014variational}, the quantum approximate optimization algorithm (QAOA) \cite{farhi2014quantum}, 
and all variational implementations of quantum machine learning algorithms \cite{schuld2020circuit, benedetti2019parameterized, perdomo2018opportunities, killoran2019continuous}.
In practice, all these cases share the same workflow:  a quantum computer is used for measuring one or more expectation values which are classically combined into some cost 
function to be minimized. For example, the cost function could be the expectation value of some given observable (VQE, QAOA, etc.) or, as typical in quantum machine learning, 
some function which quantifies the distance between an ideal model and its variational approximation.
To minimize the cost function one can use different iterative methods which often require estimation of derivatives of expectation values with respect to the variational 
parameters of the circuit. The analytic gradient of a quantum circuit can be experimentally evaluated using the so called {\it parameter-shift rule} 
\cite{li2017hybrid, mitarai2018quantum, schuld2019evaluating, mitarai2019methodology}, which allows implementation of all gradient-based optimizers such as gradient descent.

This work is structured in two main parts.
In the first part, we explore a variety of derivative estimators for quantum circuits, seeking to understand and map out the effectiveness of these different methods.
Our first result is a generalization of the analytic {\it parameter-shift rule} 
to derivatives of arbitrary orders including, as particular cases, the Hessian and the Fubini-Study  metric tensor \cite{cheng2010quantum,liu2019quantum}. 
Moreover, we analyze the analytic and finite-difference methods of estimating derivatives from a statistical perspective, 
taking into account the uncertainty which is unavoidable in any quantum computation involving a finite number of measurement shots.

In the second part of this work, we apply the theoretical tools studied in the first part to implement several optimization methods which require the experimental estimation of the gradient and higher-order derivatives. Specifically we focus on the Newton optimizer and the quantum natural gradient optimizer. 
We also consider a diagonal approximation of the Newton optimizer (borrowed from classical machine learning \cite{becker1988improving}) which has a negligible overhead compared to gradient descent but, nonetheless, is sensitive to the curvature of the cost function with respect to individual parameters.

Our generalization of the parameter-shift rule is complementary to other approaches for evaluating derivatives of quantum circuits. 
In Ref.~\cite{mitarai2020theory} it was shown how to evaluate the derivatives which are necessary to implement a VQE algorithm in a quantum chemistry scenario. 
In Ref.~\cite{mitarai2019methodology} a methodology for replacing direct measurements with indirect ones was proposed, including an experimentally feasible method for estimating the metric tensor. 
Very recently, different generalizations of the parameter-shift rule to arbitrary gates  \cite{crooks2019gradients, banchi2020measuring} 
have been proposed which, in principle, could be directly applied to extend the domain of applicability of our results to arbitrary gates.
Moreover it is worthwhile to mention that the same $0$ and $\pm \pi/2$ shifts which naturally emerge in our analysis were also used in 
Ref.~\cite{ostaszewski2019quantum} for a direct minimization of the cost function with respect to single parameters.

This work is structured as follows. In Sec.~\ref{sec:setting} 
we set the notation and define the derivative tensor. In Sec.~\ref{sec:psr} we derive the higher-order parameter-shift rules. 
In Sec.~\ref{sec:estimators} the statistical noise associated to analytic and finite-difference estimators is studied. 
In Sec.~\ref{sec:optimizers} the implementation of second-order optimizers is considered within the context of parameter-shift rules. 
Section~\ref{Sec:Numerics} focuses on investigating our findings in simulation and on
hardware. We first give practical examples of gradient and Hessian estimation and then
we use these quantities to minimize the expectation value of a simple variational circuit\footnote{Source code for the numerics and experiments in this paper is available at \url{https://github.com/XanaduAI/derivatives-of-variational-circuits}.}.

\section{Setting}
\label{sec:setting}
Most algorithms designed for near-term quantum computers are based on the
variational optimization of quantum circuits 
\cite{peruzzo2014variational, schuld2020circuit, perdomo2018opportunities, mcclean2016theory, sim2019expressibility, killoran2019continuous}. 
A variational quantum circuit acting on $n$
qubits is a sequence of gates depending on a set of classical parameters $\bm \theta =
(\theta_1, \theta_2, \dots, \theta_m)$ and producing a family of unitary operations $U(\bm
\theta)$. Typically, the circuit is applied to a fixed reference state 
$|0\rangle$ and the expectation value of a Hermitian observable $M$ is measured:

\begin{align} \label{expval}
f(\bm \theta) = \langle 0 | U(\bm \theta)^\dag M U(\bm \theta) | 0 \rangle.
\end{align}

%The previous quantity is the ideal theoretical value. In a real quantum computation with 
%$N$ measurement shots, we can only estimate $f(\bm \theta)$ up to a finite statistical 
%uncertainty, i.e., what we can actually measure is:
%\begin{align}\label{est_expval}
%\hat f(\bm \theta) = f(\bm \theta) + \hat \epsilon,
%\end{align}
%where $\hat \epsilon$ is a zero-mean random variable with variance $\sigma = %\sigma_0 / N$, 
%and $\sigma_0$ is the variance for a single shot which depends on specific details of the 
%circuit and of the observable.

The expectation value given in Eq.~\eqref{expval} (or some cost 
function which depends on it) is usually optimized with respect to the parameters $\bm 
\theta$. For example, if $M$ is the Hamiltonian of a system, we can approximate the 
ground state energy by minimizing $f(\bm \theta)$. 
Many optimization methods require to estimate the gradient
 at each iteration. For more advanced optimizers (e.g., Newton's method), one also needs the
Hessian matrix or higher-order
derivatives.

In this work we are interested in evaluating derivatives of an arbitrary
order $d$, which we can cast as a tensor with $d$ indices $j_1, j_2, \dots, j_d$ whose
elements are the real quantities

\begin{align}
g_{j_1, j_2, \dots, j_d}(\bm \theta)= 
\frac{\partial ^ d f(\bm \theta)}{\partial \theta_{j_1}\partial \theta_{j_2} 
\dots \partial \theta_{j_d}}, \label{eq:g_tensor}
\end{align}
where the gradient and the Hessian correspond to the particular cases with $d=1$ and $d=2$, respectively.

%Without making any assumption on the structure of $U(\bm \theta)$, Eq.~\eqref{eq:g_tensor} cannot be analytically simplified. 
%On an experimental level, this implies that derivatives can only be estimated using finite-difference approximations. 
%For the finite-difference method, one has to run the same circuit with small shifts of the parameters and take a linear combination of the measurement results. 

%In the next section we show how to analytically evaluate Eq.~\eqref{eq:g_tensor} in terms of 
%simple exact formulas (parameter-shift rules), for a specific but commonly-used class of variational quantum circuits.

A typical structure of many variational circuits which can be executed by near-term quantum computers is the following:

\begin{align}\label{var_circ}
U(\bm \theta) = V_m U_m(\theta_m) \dots V_2 U_2(\theta)\, V_1 U_1 (\theta_1),
\end{align}
where $V_j$ are constant arbitrary circuits, while $U_j(\theta_j)$ are ``rotation-like" 
gates, i.e., characterized by a generator $H_j$ such that $H_j^2=\mathbbm 1$ (involutory matrix) and so:

\begin{align}\label{rot_like}
U_j(\theta_j) = e^{- \frac{i}{2} H_j \theta_j} = \cos(\theta_j/2) \mathbbm 1 - i 
\sin(\theta_j/2) H_j.
\end{align}
For example, all single-qubit rotations belong to this class. 
More generally, $H_j$ can be any multi-qubit tensor product of Pauli matrices.

Note that this class can be extended to general gates. A possible method is to
decompose a gate into a product of rotation-like gates \cite{crooks2019gradients}. Another 
approach is to decompose an arbitrary generator $H_j$ into a linear combination of Pauli
operators and then evaluating derivatives with respect to each term using the stochastic
method recently proposed in Ref.~\cite{banchi2020measuring}.
\section{Parameter-shift rules}
\label{sec:psr}

The class of variational quantum circuits specified by
Eqs.~\eqref{var_circ} and~\eqref{rot_like} are commonly used
and there exists a simple parameter-shift rule to evaluate their
gradients~\cite{li2017hybrid, mitarai2018quantum, schuld2019evaluating, mitarai2019methodology}. This rule
provides the gradient analytically by evaluating the circuit
with fixed shifts of $\pi / 2$ in the parameters $\bm{\theta}$.

We begin this section by showing how the established gradient
rule can be generalized to arbitrary parameter shifts and then
extend our findings to higher order derivatives such as the
Hessian and Fubini-Study metric tensor.

\subsection{First-order derivatives: the gradient} 
 
From the previous identity, the unitary conjugation of an arbitrary operator $\hat K$
by $U_j(\theta_j)$ can always be reduced to the sum of three terms:
\begin{align}\label{K_conj}
\hat K(\theta_j)=U_j(\theta_j)^\dag \hat K U_j(\theta_j) = \hat A + \hat B \cos(\theta_j) + 
\hat C \sin(\theta_j),
\end{align}
where $\hat A, \hat B, \hat C$ are operators independent of $\theta_j$ and involving only 
$\hat K$ and $\hat H_j$. Moreover, from the standard trigonometric addition and subtraction 
identities, we can deduce that:

\begin{align}
\frac{d \cos(x)}{dx} = \frac{ \cos(x + s) - \cos(x - s)}{2 \sin(s)},  \label{d_cos}\\
\frac{d \sin(x)}{dx} = \frac{ \sin(x + s) - \sin(x - s)}{2 \sin(s)}, \label{d_sin}
\end{align}
which are valid for any $s \neq k \pi$, $k \in \mathbb Z$. Differentiating both sides of 
Eq.~\eqref{K_conj} and using Eqs.~\eqref{d_cos} and 
\eqref{d_sin}, we obtain a parameter-shift rule which is valid at the operator level:

\begin{align}\label{K_partial_shift}
\frac{\partial }{\partial{\theta_j}} \hat K(\theta_j)
= \frac{\hat K(\theta_j + s) - \hat K(\theta_j - s)}{2 \sin(s)}.
\end{align}
Note that, even if the previous expression looks like a finite-difference approximation, it 
is actually exact and we can use it to analytically estimate the gradient of expectation 
values whenever the rotation-like property of Eq.~\eqref{rot_like} holds.
Indeed the operator identity in Eq.~\eqref{K_partial_shift} can be directly applied into Eq.~\eqref{expval} 
to evaluate the $j_{\rm th}$ component of the gradient. The result 
is a family of {\it parameter-shift rules}:
\begin{align}\label{ps_exact}
g_j(\bm \theta)  = 
\frac{f(\bm \theta + s \e_j) - f(\bm \theta- s \e_j)}{2 \sin(s)}
\end{align}
where $\e_j$ is the unit vector along the $\theta_j$ axis.

Note that, in the limit $s\rightarrow 0$, $\sin(s)$ can be approximated by $s$ and we recover the 
central-difference approximation for the first derivative. On the other hand, for $s=\pi/2$ we obtain the 
parameter-shift rule already studied
in \cite{li2017hybrid, mitarai2018quantum, schuld2019evaluating, mitarai2019methodology} and used in 
quantum software libraries \cite{bergholm2018pennylane, broughton2020tensorflow, qulacs, tequila, luo2019yao}. It is important to remark that 
the formula in Eq.~\eqref{ps_exact} is exact for any choice of $s$. Strictly speaking $s$ cannot be a multiple of 
$\pi$ because of the diverging denominator, however, all the corresponding discontinuities are removable.

\subsection{Second-order derivatives: the Hessian}\label{Sec:Hessian}

A useful property of Eq.~\eqref{ps_exact} is that it can be 
iterated to get higher-order derivatives. Applying the same rule twice we get a similar formula for the Hessian:

\begin{align}\label{hess_ps_exact_s1s2}
g_{j_1, j_2}(\bm \theta)  
=& [f(\bm \theta + s_1 \e_{j_1}+ s_2 \e_{j_2}) - f(\bm \theta - s_1 \e_{j_1}+ s_2 \e_{j_2}) \nonumber  \\
& -f(\bm \theta + s_1 \e_{j_1}- s_2 \e_{j_2}) + f(\bm \theta - s_1 \e_{j_1}- s_2 \e_{j_2})] \nonumber \\
& /[4 \sin(s_1)\sin(s_2)]
\end{align}
which, for $s1=s2=s$, simplifies to:

\begin{align}\label{hess_ps_exact_s}
g_{j_1, j_2}(\bm \theta)  
=& [f(\bm \theta + s (\e_{j_1} + \e_{j_2})) - f(\bm \theta + s (-\e_{j_1} + \e_{j_2}))  \nonumber  \\
& -f(\bm \theta + s (\e_{j_1} - \e_{j_2}))+ f(\bm \theta - s (\e_{j_1} + \e_{j_2}))] \nonumber \\
& /[2 \sin(s)]^2.
\end{align}

Also in this case, for $s\rightarrow 0$, we get the standard central-difference formula for the Hessian.
For $s=\pi/2$, we get an analytic parameter-shift rule which is similar to the gradient formula used in
 Refs.~\cite{li2017hybrid, mitarai2018quantum, schuld2019evaluating, mitarai2019methodology}, but extended 
 to the Hessian. A formula equivalent to Eq.~\eqref{hess_ps_exact_s} for the particular case $s=\pi/2$ was recently used in Ref.~\cite{huembeli2020characterizing} and  
 Ref.~\cite{mitarai2020theory}.
Particular attention should be paid to the diagonal of the Hessian since, for each element, two shifts 
are applied to the same parameter $\theta_j$. In this case, two alternative choices for the value of $s$ 
which appears in Eq.~\eqref{hess_ps_exact_s} are particularly relevant. For the choice $s=\pi/2$, we get:
\begin{align}\label{hess_diag_pi}
g_{j, j}(\bm \theta)  
=& [f(\bm \theta + \pi \e_{j}) - f(\bm \theta)]/2,
\end{align}
where we used that $f(\bm \theta + \pi \e_{j}) = f(\bm \theta - \pi \e_{j})$.
Instead, for $s=\pi/4$, we obtain:
\begin{align}\label{hess_diag_half}
g_{j, j}(\bm \theta)  
=& [f(\bm \theta + \e_{j} \pi/2 ) - 2 f(\bm \theta) + f(\bm \theta - \e_{j} \pi/2)]/2.
\end{align}
Each of the two previous formulas has alternative advantages. The advantage of Eq.~\eqref{hess_diag_pi} is
that it involves only two expectation values, and so it is more direct with respect to
Eq.~\eqref{hess_diag_half}, which is instead a linear combination of three terms.
On the other hand, the parameter shifts involved in Eq.~\eqref{hess_diag_half} are only $\pm\pi/2$. 
This implies that all the elements of the full Hessian matrix can be evaluated using only the same 
type of $\pm\pi/2$ shifts and this fact could be an experimentally relevant simplification. 
Moreover, in the typical scenario in which one has already evaluated the gradient using the $m$ 
pairs of shifts $f(\bm \theta \pm \pi/2 \e_{j})$, Eq.~\eqref{hess_diag_half} allows to evaluate 
the diagonal of the Hessian with the extra cost of just a single expectation value (i.e., $f(\bm \theta)$). 
In Section \ref{sec:optimizers} we show how this fact can be conveniently exploited to replace 
the vanilla gradient descent optimizer with a diagonal approximation of the Newton optimizer, with a
negligible computational overhead.

\subsection{Fubini-Study metric tensor}
\label{sec:metric-tensor}
A second-order tensor which plays an important role in quantum information theory is the Fubini-Study metric tensor which, 
for a pure variational state $|\psi(\bm\theta)\rangle$, can be expressed as:
\begin{align}\label{metric-tensor}
F_{j_1,j_2}(\bm \theta)= - \frac{1}{2}\left.
\frac{\partial^2}{\partial \theta_{j_1}\partial \theta_{j_2}}  |\langle \psi(\bm \theta')|\psi(\bm \theta)\rangle|^2 \right|_{\theta'=\theta},
\end{align}
and corresponds to the real part of the quantum geometric tensor (see e.g., \ Appendix A1 of \cite{stokes2020quantum} for a detailed derivation). 
For pure states and up to constant factors, Eq.~\eqref{metric-tensor} can also be associated to other tensors such as the quantum 
Fisher information matrix or the Bures metric tensor \cite{liu2019quantum}. 
Since in this work we only deal with pure states, 
we often refer to Eq.~\eqref{metric-tensor} simply as ``the metric tensor''.

Differently from the Hessian, the metric tensor is not linked to a particular observable $M$ but it 
is instead a geometric property of a variational quantum state, which in our setting is simply $|\psi(\bm \theta)\rangle = U(\bm \theta) |0\rangle$. 
This tensor plays a crucial role in the implementation of the quantum natural gradient optimizer \cite{stokes2020quantum} and 
in the variational quantum simulation of imaginary-time evolution \cite{mcardle2019variational}.

Now we can make a useful observation: the metric tensor in Eq.~\eqref{metric-tensor} can actually be seen (up to a constant factor) as the Hessian 
of the expectation value $f(\bm \theta)$ defined in equation Eq.~\eqref{expval}, for the particular 
observable $M(\bm \theta')=U(\bm \theta') |0\rangle \langle 0 | U(\bm \theta')$.
Therefore all the previous
theoretical machinery that we have derived for the Hessian applies also to the metric tensor and we 
get the corresponding  parameter-shift rule which is simply the same as Eq.~\eqref{hess_ps_exact_s} 
where each expectation value is:
\begin{align}\label{survival_prob}
f(\bm \theta) = |\langle \psi(\bm \theta')|\psi(\bm \theta)\rangle|^2.
\end{align}
 
The quantity  $|\langle \psi(\bm \theta')|\psi(\bm \theta)\rangle|^2$ is the survival probability 
of the state $|0\rangle$ after the application of the
circuit $U(\bm \theta')U(\bm \theta)$. This probability can be easily estimated with near-term 
quantum computers either with a swap-test, or more simply, as the probability of obtaining the $00\dots 0$ 
bit string after measuring the state $U(\bm \theta')U(\bm \theta) |0\rangle$ in the computational basis.

Replacing Eq.~\eqref{survival_prob} into Eq.~\eqref{hess_ps_exact_s}, setting $s=\pi/2$ and $\theta'=\theta$,  
we get the explicit parameter-shift rule for the metric tensor:
\begin{align} \label{metric_rule}
F_{j_1, j_2}(\bm \theta) =
& -\frac{1}{8} \Big [|\langle \psi(\bm \theta)|\psi(\bm \theta + (\e_{j_1} + \e_{j_2}) \pi/2)\rangle|^2 \nonumber \\
&-|\langle \psi(\bm \theta)|\psi(\bm \theta + (\e_{j_1} - \e_{j_2}) \pi/2)\rangle|^2     				\nonumber \\
&-|\langle \psi(\bm \theta)|\psi(\bm \theta + (-\e_{j_1} + \e_{j_2}) \pi/2)\rangle|^2     				\nonumber \\
&+|\langle \psi(\bm \theta)|\psi(\bm \theta - (\e_{j_1} + \e_{j_2}) \pi/2)\rangle|^2  \Big].
\end{align}

As we discussed for the Hessian, also in this case the formula for the diagonal elements can be simplified in two alternative ways. The first
formula, corresponding to Eq.~\eqref{hess_diag_pi}, is:
\begin{align}\label{metric_rule_diag_pi}
F_{j, j}(\bm \theta) =
& \frac{1}{4}\Big [1 - |\langle \psi(\bm \theta)|\psi(\bm \theta + \pi \e_{j})\rangle|^2 \Big].
\end{align}
The second equivalent formula, corresponding to  Eq.~\eqref{hess_diag_half}, is:
\begin{align}\label{metric_rule_diag_half}
F_{j, j}(\bm \theta) =
&\frac{1}{2}\Big [ 1 - |\langle \psi(\bm \theta)|\psi(\bm \theta + \e_{j}  \pi /2)\rangle|^2\Big] \,,
\end{align}
where we used that $|\langle \psi(\bm \theta)|\psi(\bm \theta + \e_{j}  \pi /2)\rangle|^2 = |\langle \psi(\bm \theta)|\psi(\bm \theta - \e_{j}  \pi /2)\rangle|^2$.
We also comment that an efficient method for evaluating diagonal blocks of the metric tensor was proposed 
in \cite{stokes2020quantum}. Moreover, an experimentally feasible methodology for measuring the metric tensor was 
proposed in \cite{mitarai2019methodology}.

\subsection{Arbitrary-order derivatives} 
The same iterative approach can be used to evaluate derivatives of arbitrary order.
In this case, for simplicity, we set $s=\pi/2$ and we introduce the multi-parameter shift
vectors:

\begin{align}\label{multi_shift}
 {\bf k}_{\pm j_1, \pm j_2, \dots, \pm j_d} =\frac{\pi}{2} (\pm \e_{j_1} \pm \e_{j_2} \dots \pm \e_{j_d}),
\end{align}
where $j_1, j_2, \dots, j_d$ are the same $d$ indices which appear also in the derivative tensor defined in Eq.~\eqref{eq:g_tensor}.
These vectors represent all the shifts in parameter space that are generated by iterating the parameter-shift rule 
of Eq.~\eqref{ps_exact} for $j \in \{j_1, j_2, \dots, j_d\}$. We explicitly introduce the set of all shifts which are
related to a derivative of order $d$:

\begin{align}\label{S_multi_shift}
 S_{j_1, j_2, \dots, j_d} = \{ {\bf k}_{\pm j_1, \pm j_2, \dots, \pm j_d} , \; \forall \text{ choices of signs}\}.
\end{align}
For example, for the Hessian formula given in Eq.~\eqref{hess_ps_exact_s} evaluated as $s=\pi/2$ there are $4$ possible shifts:
$S_{j_1, j_2} = \{{\bf k}_{\pm j_1, \pm j_2}\}$.

With this notation, the derivative tensor of order $d$ defined in Eq.~\eqref{eq:g_tensor}, can be expressed 
in terms of the following generalized  parameter-shift rule:

\begin{align}\label{d_rule}
g_{j_1, j_2, \dots, j_d}(\bm \theta)=\frac{1}{2^d} \sum_{{\bf k} \in S_{j_1,j_2, \dots, j_d}} \mathcal P ({\bf k})\, f(\bm \theta + {\bf k}).
\end{align}
where $\mathcal P({\bf k})$ is the parity of a shift vector, i.e., the parity of negative indices in Eq.~\eqref{multi_shift}.

How many expectation values are necessary to evaluate the $d$-order derivative defined in Eq.~\eqref{eq:g_tensor}?
This is given by the number of elements in the previous sum, which is  $|S_{j_1, j_2, \dots, j_d}|=2^d$.
When some of the indices are repeated, the number of shifts can be further reduced but we neglect this fact for the moment.
Taking into account that the derivative tensor is symmetric with respect to permutations of the indices, the number of distinct elements of a tensor of order $d$ is given by the combinations of $d$ indices sampled with replacement from the set of $m$ variational parameters, corresponding to the multiset coefficient $\binom{m + d - 1}{d}$ .
Therefore, the total number of expectation values to evaluate
a derivative tensor of order $d$ is bounded by:
\begin{align}\label{num_exp_vals_bound}
\# \text{ expectation values} \le 2^d \binom{m + d - 1}{d} = \mathcal O(m^d),
\end{align}
where in the last step we assumed $m >> d$. In the opposite regime $d >> m$, each angle parameter $\theta_j$ can be subject to multiple shifts. 
However, because of the periodicity of $f(\bm \theta)$, for each $\theta_j$ we have at most $4$ shifts: $0, \pm \pi/2, \pi$. 
Moreover, from Eq.~\eqref{K_conj} it is easy to check that a $\pi$ shift can be expressed as a linear combination of the others:

\begin{align}\label{pi-shift}
f(\bm \theta \pm \pi \e_{j}) = f(\bm \theta + \e_{j} \pi/2 ) + f(\bm \theta -\e_{j}  \pi/2 ) - f(\bm \theta),
\end{align}
which means that only 3 shifts for each parameter are actually enough: the trivial $0$ shift and the two $\pm \pi/2$ shifts.

From this simple counting argument we get another upper bound:
\begin{align}\label{num_exp_vals_bound2}
\# \text{ expectation values} \le 3^m,
\end{align}
which is valid for any $d$.
This implies an interesting fact: from a finite number of $3^m$ combinations of variational parameters $\bm \theta$ we 
can Taylor-expand $f(\bm \theta)$ up to an arbitrary order and so we can actually evaluate $f(\bm \theta)$ for all 
possible values of $\bm \theta$.

This fact may seem quite surprising, however, it simply reflects the trigonometric structure of rotation-like gates.
Indeed from Eq.~\eqref{K_conj}, we see that $\hat K(\theta_j)$ is a linear combination of $3$ terms. Similarly, a full 
circuit with $m$ parameters can be expressed as a linear combination of $3^m$ operators. Therefore, a generic expectation 
value $f(\bm \theta)$ is a linear combination of different trigonometric functions weighted by $3^m$ scalar coefficients, 
a fact which was already noticed in Ref.~\cite{ostaszewski2019quantum}. In principle, these coefficients could be fully 
determined by evaluating $f(\bm \theta)$ at $3^m$ different values of $\bm \theta$.
In this work we focused on the different combinations of the 3 specific shifts of $0, \pm \pi/2$ for each angle $\theta_j$. 
However, one could also obtain similar results using any other choice of 3 non-trivial shifts for each angle.

This concludes the first part of this work, in which we derived several exact analytical results. 
However, if we really want to apply Eqs.~\eqref{ps_exact}, 
\eqref{hess_ps_exact_s} and \eqref{d_rule} in a quantum  experiment, we have to take into account that each 
expectation value on the r.h.s of these equations can be measured only up to a finite precision and so the derivative 
on the l.h.s. can be estimated only up to a finite error. 

\section{Statistical estimation of derivatives}\label{Sec:FiniteShots}
\label{sec:estimators}

The expectation value in Eq.~\eqref{expval} is a theoretical quantity. In a real quantum computation with 
$N$ measurement shots, we can only estimate $f(\bm \theta)$ up to a finite statistical 
uncertainty, i.e., what we actually measure is
\begin{align}\label{est_expval}
\hat f(\bm \theta) = f(\bm \theta) + \hat \epsilon,
\end{align}
where $\hat \epsilon$ is a zero-mean random variable with variance $\sigma^2 = \sigma_0^2 / N$, 
and $\sigma_0^2$ is the variance for a single shot which depends on the specific details of the 
circuit and of the observable.
The aim of this section is to take into account the statistical noise which appears in Eq.~\eqref{est_expval}. 
We consider this problem within the theoretical framework of statistical inference and we study different {\it derivative estimators} 
for the quantities defined in Eq.~\eqref{eq:g_tensor}. In particular we are interested in two main classes of estimators: 
one based on a finite-shot version of the analytic parameter-shift rules derived in the previous section and one 
based on a standard finite-difference approximation.

More precisely the analytic estimator for a derivative tensor of arbitrary order $d$ is:

\begin{align}\label{d_rule_est}
\hat g_{j_1, j_2, \dots, j_d}^{(s=\pi/2)}=\frac{1}{2^d} \sum_{{\bf k} \in S_{j_1,j_2, \dots, j_d}} \mathcal P ({\bf k})\, \hat f(\bm \theta + {\bf k}),
\end{align}
where we simply placed Eq.~\eqref{est_expval} into \eqref{d_rule}.

We can already make a preliminary comparison by considering the central-difference approximation which has 
a very similar structure but involves an infinitesimal step size $h$ instead of $s=\pi/2$. 
In order to simplify the comparison we express it using the same notation introduced in Eqs.~\eqref{multi_shift} and \eqref{S_multi_shift}:
\begin{align}\label{d_fd_est}
\hat g_{j_1, j_2, \dots, j_d}^{(h)}=\frac{1}{(2h)^d} 
\sum_{{\bf k} \in S_{j_1,j_2, \dots, j_d}} \mathcal P ({\bf k})\, \hat f(\bm \theta + {\bf k}\, 2h/\pi).
\end{align}
It is evident that the only difference introduced in Eq.~\eqref{d_fd_est}, compared to \eqref{d_rule_est}, 
is that shifts have a step size $h$ and that the full estimator is scaled by $h^{-d}$. 
This directly implies that, for $h<1$, the statistical variance of the central-difference estimator is 
amplified by a factor of $h^{-2d}$, which is a strong limitation with respect to the analytic estimator. 
A more detailed analysis of the statistical error characterizing the analytic and the 
finite-difference methods will be presented later for the particular case of the gradient ($d=1$).

\subsection{Quantifying the error of a statistical estimator}

In general terms, our aim is to find a good estimator $\hat g_{j_1,\dots, j_d}(\bm \theta)$ which only 
depends on a finite number of measurement shots and which should approximate the derivative tensor 
$g_{j_1,\dots, j_d}(\bm \theta)$ defined in Eq.~\eqref{eq:g_tensor} well. As a figure of merit for 
the performance of an estimator we can use its {\it mean squared error} (MSE) with 
respect to the true value which is
\begin{align}\label{mse}
\Delta(\hat g_{j_1,\dots, j_d}) = \mathbbm E [(\hat g_{j_1,\dots, j_d} - g_{j_1,\dots, j_d})^2 ],
\end{align}
where $\mathbbm E (\cdot)$ is the (classical) average over the statistical distribution of 
the measurement outcomes. 
The performance of the estimator can also be measured with respect to the full tensor in terms of the total error,
\begin{align}\label{mse_j}
\Delta ( \hat{\bm g}) = \mathbbm E \left [ \|\hat {\bm g} - \bm g \|^2 \right] 
= \sum_{j_1,\dots, j_d} \Delta(\hat g_{j_1,\dots, j_d}),
\end{align}
which is simply the sum of the single-element errors.

Before considering specific derivative estimators, it is useful to remind also about the notions 
of {\it bias} and {\it variance}:
\begin{align}
{\rm Bias} ( \hat g_{j_1,\dots, j_d}) &
:= \mathbbm E(\hat g_{j_1,\dots, j_d}) - g_{j_1,\dots, j_d} \label{bias} \\
{\rm Var} ( \hat g_{j_1,\dots, j_d}) &
:= \mathbbm E(\hat g_{j_1,\dots, j_d}^2) - \mathbbm E(\hat g_{j_1,\dots, j_d})^2. \label{var}
\end{align}
These two quantities correspond to different errors: the bias represents a constant error 
which remains present even in the limit of an infinite number of shots $N \rightarrow 
\infty$, while the variance represents statistical fluctuations which are due to a finite 
$N$. 
It is well known that both effects can increase the MSE, and indeed we have
\begin{align}\label{mse_bias_var}
\Delta(\hat g_{j_1,\dots, j_d}) = {\rm Var} ( \hat g_{j_1,\dots, j_d}) + {\rm Bias} ( \hat g_{j_1,\dots, j_d})^2.
\end{align}
In the next subsections we focus on the estimation of the gradient, but a similar analysis can be
extended to the Hessian and higher-order derivatives.

\subsection{Finite-difference gradient estimator}\label{Sec:FDEstimator}

A general way of approximating the gradient is to use a  finite-difference estimator, which 
requires to experimentally measure expectation values for slightly different values of the parameters. 
The most common finite-difference estimators are the {\it forward-difference} and the {\it central-difference}, 
both well studied in classical numerical analysis \cite{gill2019practical}. In this work we focus mainly on 
the central-difference estimator because it has the same structure of to the parameter-shift rules derived 
in the previous section and so we can easily make a comparison.

Given a fixed step size $h>0$, the symmetric {\it finite-difference estimator} for the 
$j_{\rm th}$ element of the gradient can be defined as:
\begin{align}\label{fd_est}
\hat g_j^{(h)} &= \frac{\hat f(\theta_j + h) - \hat f(\theta_j - h) }{2h}  \nonumber\\
&=\frac{f(\theta_j 
+ h) - f(\theta_j - h) }{2h} + \frac{\hat \epsilon_+ - \hat \epsilon_- }{2h},
\end{align}
where we used Eq.~\eqref{est_expval} and $\hat \epsilon_{\pm}$ is the statistical noise
associated to $\hat f(\theta_j \pm h)$.
In the right-hand-side of the previous equation we have the sum of two terms which are the 
finite-difference approximation and the statistical noise, respectively. Each term introduces
a different kind of error: one is linked to finite step size $h$, the other is linked to the 
finite number of shots $N$. Such errors correspond to the bias and to the variance of the 
estimator discussed in the previous subsection. Indeed, from Eqs.~\eqref{bias} and 
\eqref{var}, we have:
\begin{align}
{\rm Bias} ( \hat{g}_j^{(h)}) 
&= \frac{f(\theta_j + h) - f(\theta_j - h) }{2h} - g_j \nonumber \\
&= \frac{f_3 h^2}{3!} + O(h^3), 
\label{bias_h} \\
{\rm Var} ( \hat{g}_j^{(h)}) 
&= \frac{\sigma^{2}_{0}(\theta_j + h) + \sigma^{2}_{0}(\theta_j - h)}{4 N h^2} \label{intermediate_var_h} \\ % \nonumber \\
&\approx \frac{\sigma_0^{2}}{2 N h^2}, \label{var_h}
\end{align}
where, in Eq.~\eqref{bias_h}, we used 
the Taylor-series approximation with $f_3 = \partial^3 f(\theta_j)/\partial \theta_j^3$, and
$\sigma_{0}^{2}(\theta_{j} \pm h)$ is the single-shot variance evaluated at
$\theta_{j} \pm h$.
The step from Eq.~\eqref{intermediate_var_h} to Eq.~\eqref{var_h} is justified only if the following assumption holds.\\

\noindent{\bf Assumption 1:} The variance of the measured observable depends weakly on the parameter shift, such that $\sigma^{2}_{0}(\theta_j + x) + \sigma^{2}_{0}(\theta_j - x) \simeq 2 \sigma^{2}_{0}$ for any value of $x$.
\\

The previous assumption is usually a quite good approximation, however one can easily construct counter-examples 
in which this is violated for large values of the shift. For this reason, whenever a subsequent result 
depends on Assumption 1, it will be explicitly stressed.

It is evident that the terms in Eqs.~\eqref{bias_h} and \eqref{var_h}   have opposite scaling with respect to the step size: 
for small $h$ the variance diverges, while for large $h$ the bias dominates. This trade-off 
implies that there must exist an optimal choice of $h$.
%, an effect which is confirmed by our numerical experiments (see Fig. \ref{fig:fd_vs_h}).

Substituting   Eqs.~\eqref{bias_h} and \eqref{var_h} into Eq.~\eqref{mse_bias_var} we get, 
up to $\mathcal O(h^{6})$ corrections and within the validity of Assumption 1, the MSE for an arbitrary step size:
\begin{align}\label{mse_h}
\Delta(\hat g_j^{(h)}) \simeq \frac{\sigma_0^{2}}{2 N h^2} + \frac{f_3^2 h^4}{36}.
\end{align}
Imposing the derivative with respect to $h$ to be zero and assuming $f_3\neq 0$, 
we get the optimal step size $h^*$ and the optimal error:
\begin{align}
h^* &= \left( \frac{9 \sigma_0^{2}}{ f_3^2 N} \right)^{\frac{1}{6}} \propto N^{-\frac{1}{6}} 
\simeq N^{-0.167}, \label{h_opt} \\
\Delta(\hat g_j^{(h^*)}) &=\frac{3}{2} \frac{\sigma_0^{2}}{2N} (h^*)^{-2} =  \frac{3}{2} \left[\frac{\sigma_0^{2}}{2 N}\right]^{2/3} 
\left[\frac{f_3^2}{18}\right]^{1/3}  \nonumber \\
&\propto N^{-2/3} \simeq N^{-0.667}, \label{delta_h_opt}
\end{align}
which are valid only for sufficiently large $N$ (i.e., for sufficiently small $h^*$).
When the number 
of shots is small, the optimal step $h^*$ can become so large that both the Taylor 
approximation and Assumption 1 may not be valid anymore, so that we could observe deviations from the predictions 
of Eqs.~\eqref{mse_h}, \eqref{h_opt} and \eqref{delta_h_opt}. 

From a practical point of view, it is not straightforward to determine the optimal step $h^*$ since 
it depends on the third derivative $f_3$. The situation is significantly simpler for rotation-like gates where, 
from Eq. \eqref{K_conj}, we have that the third derivative is equal to the first and so $f_3$ can be 
replaced by $g_j$. However, as we are going to see in the next subsections, in this case it is more convenient to use the parameter-shift estimator.

For the sake of completeness we comment that, by repeating the same analysis for the forward difference 
estimator $\hat G_j^{(h)}=[\hat f(\theta_j + h) - \hat f(\theta_j)]/h$, one gets similar scaling laws but with different exponents:
\begin{align} \label{mse_h_forward}
\Delta(\hat G_j^{(h)}) &\simeq \frac{\sigma_0^{2}}{2 N h^2} + \frac{f_2^{\,2} h^2}{4} ,\\
h^* &= \left(\frac{2\sigma_0^2 }{f_2^2 N}\right)^{\frac{1}{4}}\propto N^{-0.25}, \label{h_opt_forward} \\
\Delta(\hat G_j^{(h^*)}) &= \frac{\sigma_0^{2}}{N} (h^*)^{-2} =  
\left (\frac{\sigma_0^{2} f_2^{\,2}}{2N}\right)^{\frac{1}{2}}\propto N^{-0.5}\;, \label{delta_h_opt_forward}
\end{align}
which are valid as long as the approximations in the Taylor truncation and in Assumption 1 are justified.

It is notable that a similar trade-off for the choice of the step-size $h$ was studied 
also in the field of classical numerical analysis (see e.g., \cite{gill2019practical} \cite{mathur2012analytical}). In a 
classical computer the error term $\hat \epsilon$ in Eqs.~\eqref{est_expval} and 
\eqref{fd_est} is the {\it round-off error} (or {\it machine epsilon}) associated with the 
finite-precision representation of real numbers. Optimizing the value of $h$ was particularly 
useful in the early days of classical computing, because of the limited memory and 
computational resources of that time. Nowadays, with current quantum computers, we are in a 
similar situation: the statistical uncertainty of quantum measurements and the limited 
number measurement shots give rise to a {\it quantum round-off error}, which is usually 
several orders of magnitude larger than the classical counterpart. For this reason, it is 
likely that many old problems of classical numerical analysis will become relevant again 
in the context of quantum computing.

\subsection{Parameter-shift gradient estimators}

In Section \ref{sec:psr} we derived, for a large class of variational circuits, an exact formula for the gradient 
which is given in Eq.~\eqref{ps_exact}.
For a finite number of shots, we can define the corresponding {\it parameter-shift estimator}:

\begin{align}\label{ps_est}
\hat g_j^{(s)} = \frac{\hat f(\theta_j + s) - \hat f(\theta_j - s)}{2 \sin(s)} 
= g_j + \frac{\hat \epsilon_+ - \hat \epsilon_-}{2 \sin(s)},
\end{align}
where $\hat \epsilon_\pm$ is the statistical noise associated to the measurement of 
$\hat f(\theta_j \pm s)$.
Differently from the finite-difference estimator $\hat g_j^{(h)}$ presented in 
Eq.~\eqref{fd_est}, $\hat g_j^{(s)}$ is unbiased because in this case there is no finite-step 
error since Eq.~\eqref{ps_exact} is exact. Specifically, we have:
\begin{align}
{\rm Bias} ( \hat{g}_j^{(s)}) 
&= 0, \label{bias_s} \\
{\rm Var} ( \hat{g}_j^{(s)}) 
&=  \frac{\sigma^{2}_{0}(\theta_j + s) + \sigma^{2}_{0}(\theta_j - s)}{4 N\sin(s)^2} \label{var_s_intermediate} \\
&\approx \frac{\sigma_0^2}{2N \sin(s)^2}. \label{var_s}
\end{align}
The step from Eq.~\eqref{var_s_intermediate} to Eq.~\eqref{var_s} is justified only if Assumption 1 is valid.

The MSE of the partial-shift estimator is only due to the statistical 
noise and, if Assumption 1 applies, this is approximated by:
\begin{align}\label{mse_s}
\Delta(\hat g_j^{(s)}) = {\rm Var} ( \hat{g}_j^{(s)})\approx\frac{\sigma_0^2}{2 N \sin(s)^2},
\end{align}

\subsubsection{The parameter-shift rule with maximum shift ($s=\pi/2$)}

The simple expression for the MSE (Eq.~\eqref{mse_s}) implies that, under the validity of Assumption 1, the 
optimal shift $s^*$ is the one which maximizes the denominator of 
Eq.~\eqref{mse_s}, i.e., $s^* = \pi/2$. This corresponds to the parameter-shift rule already studied in the 
literature \cite{li2017hybrid, mitarai2018quantum, schuld2019evaluating, mitarai2019methodology}.

The explicit definition of the estimator and its MSE are:
\begin{align}
\hat g_j^{(s=\pi/2)} &= \frac{\hat f(\theta_j + \pi/2) - \hat f(\theta_j - \pi/2)}{2}  \nonumber \\
&= g_j + \frac{\hat \epsilon_+ - \hat \epsilon_-}{2 }. \label{max_shift_est} \\
\Delta(\hat g_j^{(s=\pi/2)}) &={\rm Var} ( \hat{g}_j^{(s=\pi/2)}) \nonumber \\
&= \frac{\sigma_0^2(\theta_j + \pi/2) + \sigma_0^2(\theta_j - \pi/2)}{2 N}
\label{intermediate_mse_max_s}\\
&\simeq \frac{\sigma_0^2}{2 N}. \label{mse_max_s}
\end{align}

\subsubsection{Scaled parameter-shift gradient estimator}\label{Sec:Rescaled}
A simple way of further reducing the statistical 
error below the value of Eq.~\eqref{var_s_intermediate} is to define a {\it scaled parameter-shift estimator}, 
which is the same as Eq.~\eqref{ps_est} but scaled by a multiplicative constant $\lambda \in [0,1]$:

\begin{align}
\hat g^{(\lambda,\, s)}_j &
= \lambda  \hat g^{(s)}_j 
= \lambda g_j +  \lambda \frac{\hat \epsilon_+ - \hat \epsilon_-}{2 \sin(s) }. 
\label{scaled_est}
\end{align}
The effect of the scaling is to reduce the variance by a factor of $\lambda^2$.
However it has a cost: the new estimator is not unbiased anymore. Indeed, we have:

\begin{align}
{\rm Bias} ( \hat{g}_j^{(\lambda,\, s)}) 
&= (\lambda - 1) g_j, \label{bias_lambda} \\
{\rm Var} ( \hat{g}_j^{(\lambda,\, s)}) 
&= \lambda^2 {\rm Var} ( \hat{g}^{(s)}_j ). \label{var_lambda}
\end{align}
and so, from Eq.~\eqref{mse_bias_var}, its MSE is:
\begin{align}\label{mse_lambda}
\Delta(\hat g_j^{(\lambda,\, s)}) =\lambda^2  {\rm Var} ( \hat{g}^{(s)}_j ) + (\lambda - 1)^2 g_j^2.
\end{align}
The error is quadratic with respect to $\lambda$ and is minimized by
\begin{align} \label{lambda_star}
\lambda^*= \frac{1}{1 + \frac{ {\rm Var} ( \hat{g}^{(s)}_j )}{g_j^2}},
\end{align}
corresponding to the MSE
\begin{align} \label{mse_lambda_star}
\Delta(\hat g_j^{(\lambda^*,\, s)})= \lambda^* \Delta(\hat g_j^{(\lambda=1,\, s)}) =  \lambda^*  {\rm Var} ( \hat{g}^{(s)}_j ),
\end{align}
or, equivalently, to
\begin{align} \label{mse_lambda_star2}
\Delta(\hat g_j^{(\lambda^*,\, s)})= (1 - \lambda^*) g_j^2.
\end{align}
Eq.~\eqref{mse_lambda_star} shows that the scaled estimator is always more accurate 
than the simple parameter-shift estimator of Eq.~\eqref{max_shift_est}.
The last equation \eqref{mse_lambda_star2} is also interesting since it implies that the
relative MSE of $\hat g_j^{(\lambda^*,\, s)}$ is always less than 1, for any amount of statistical noise and so
for any $N$.

We envisage that this estimator could potentially be helpful when faced with the so-called
barren plateau phenomenon \cite{mcclean2018barren}, according to which the typical magnitude of the gradient of a random
circuit decays exponentially with respect to the number of qubits. When the gradient is much
smaller than the statistical estimation error ($ {\rm Var} ( \hat{g}^{(s)}_j ) \gg g_j$ ), the optimal scaling factor $\lambda^*$ 
can be much smaller than the vanilla value of $1$. So, in this kind of noise-dominated regime, 
the scaled estimator can be particularly advantageous if we wish to reduce the mean squared error.
Note also that from a practical point of view, since $\lambda^*$ is a constant scalar, it can be
simply absorbed into a re-normalization of the learning rate in most machine learning optimizers.
This observation could shed some light on the practical applicability of machine learning algorithms
even when the gradient is so small that its estimation is dominated by statistical noise.

\subsection{Comparison between analytic and finite-difference gradient estimators}\label{Sec:FDPSComparison}

Even if some generalization approaches have been proposed \cite{banchi2020measuring,crooks2019gradients}, it is important to stress that all the parameter-shift methods
can be directly applied only to a class of circuits (those involving involutory generators), 
while the operating regime of the finite-difference method is more general. 
However, in all those cases in which both approaches could be applied, which is the optimal one?

To give an answer to this question, it is important to look for a fair comparison and to choose 
a reasonable figure of merit.
Since the parameter-shift rule involves two symmetric shifts, it is reasonable to compare it  
to the symmetric-difference estimator and not to the forward-difference one.
As figure of merit, we chose the MSE between the estimated value and the
true value of the gradient, i.e., Eqs.~\eqref{mse} and \eqref{mse_j}.

As a first step, we are going to make a comparison between $\hat g_j^{(h)}$ and $\hat g_j^{(s)}$. 
Eventually, we will take into consideration also the scaled parameter-shift estimator $\hat g_j^{(\lambda,\, s)}$ 
defined in Eq.\eqref{scaled_est}, which will be shown to outperform the other estimators, assuming the optimal weighting parameter is known.

\subsubsection{Small step limit ($h\rightarrow 0$)}
In the small step limit $h \rightarrow 0$ ,  we notice that the finite-difference estimator 
(Eq.~\eqref{fd_est}) is approximately equal to the parameter-shift estimator 
(Eq.~\eqref{ps_est}) with $s=h$. Indeed, in this limit, $\sin(h) \simeq h$ and so

\begin{align}
\hat g_j^{(h)} \xrightarrow{h \rightarrow 0} \hat g_j^{(s = h)}.
\end{align}
We have already shown that, if the noise is independent of the variational parameter, the MSE for the 
parameter-shift estimator is minimized when $s=\pi/2$ (see Eq.~\eqref{mse_max_s}). So, in the small
$h$ limit, we have
\begin{align}
\Delta (\hat g_j^{(h)}) \xrightarrow{h \rightarrow 0} \Delta(\hat g_j^{(s = h)}) > \Delta(\hat g_j^{(s = \pi/2)}),
\end{align}
where the last inequality is valid only under the validity of Assumption 1.
In this limit, for $s=\pi/2$ we get 
a large improvement of the 
error since we have $\Delta(\hat g_j^{(h)}) / \Delta(\hat g_j^{(s = \pi/2)} ) \propto 1/h^2$.

\subsubsection{Finite step case with $h\le 1$}

For a non-infinitesimal $h\le 1$, we have a similar result. Indeed, for all $h\le 1$, there always exists an $s_h$ such that
$\sin(s_h)=h$. And so:
\begin{align}
\Delta (\hat g_j^{(h)})&= \Delta(\hat g_j^{(s=s_h)}) +
{\rm Bias}(\hat g_j^{(h)})^2   \nonumber \\
&\ge \Delta(\hat g_j^{(s=s_h)}), \label{ineq}
\end{align}
where in the first equality we used Eq.~\eqref{mse_bias_var}, the fact that $\hat g_j^{(h)}$ and $\hat g_j^{(s_h)}$ 
have equal variance, and that $\hat g_j^{(s_h)}$ is unbiased.

So, also in this case, the parameter-shift rule has a smaller error with respect to the finite-difference method. 
Furthermore, if Assumption 1 is valid, the estimator $g_j^{(s = \pi/2)}$ becomes optimal since 
$\Delta(\hat g_j^{(s=s_h)}) > \Delta (\hat g_j^{(s = \pi/2)})$.

\subsubsection{Finite step case with $h > 1$}

If $h > 1$, the problem is non-trivial. In this case, the variance of the finite-difference
estimator is smaller than the variance of the parameter-shift estimator for all values of $s$, 
i.e., ${\rm Var}(\hat g_j^{(h)})\le{\rm Var}(\hat g_j^{(s)})$. On the other hand, since $h$ is large, 
the bias of $\hat g_j^{(h)}$ can be large while $\hat g_j^{(s)}$ is unbiased. 
The trade-off between the two effects determines which method minimizes the MSE.
Moreover we should also consider that that Taylor approximation which we used for the error analysis
of the finite-difference estimator  is likely to be broken for $h>1$.

%In conclusion, while for $h\le 1$ the parameter-shift method is always optimal, if $h>1$ we have to 
%explicitly compare the two errors given in Eqs.~\eqref{mse_h} and \eqref{mse_max_s}:
%
%\begin{align}
%\frac{\Delta(\hat g^{(h)})}{\Delta(\hat g^{(s = \pi/2)}) } 
%\ge 1 \Longleftrightarrow  \frac{1}{h^2} + \frac{f_3^2 N h^4}{18 \sigma_0^2} \ge 1
%\end{align}
 
It is however intuitive that when the statistical noise is extremely large, the finite-difference formula in 
Eq.~\eqref{fd_est} for $h>1$ has a large denominator $2h$ which is able to attenuate the statistical fluctuations 
more than the parameter-shift denominator (which is equal to $2$).
So we may ask if, by simply adding a multiplicative constant $\lambda \in [0, 1]$ to the 
parameter-shift estimator we could always obtain a smaller error with respect to the finite-difference method for all values of $h$ and 
for any value of the statistical noise. This is indeed true, as we are going to show in the following subsection.
\\

\subsubsection{The scaled parameter-shift estimator is always optimal}
We have just compared the error scaling of the analytic and the finite-difference estimators.
In this subsection, we instead compare both methods with respect to the {\it scaled parameter-shift estimator} 
defined in Eq.~\eqref{scaled_est}. In particular we are going to show that $\hat g_j^{(\lambda^*,\, s)}$, 
where $\lambda^*$ is the optimal scaling parameter given in Eq.~\eqref{lambda_star}, always has a smaller 
MSE when compared to $\hat g_j^{(h)}$ or $\hat g_j^{(s)}$.

Indeed, from Eq.~\eqref{mse_lambda_star} and since $\lambda^*\le 1$, we directly have:

\begin{align}
\Delta(\hat g_j^{(s)}) \ge  \lambda^*  \Delta(\hat g_j^{(s)}) = \Delta(\hat g_j^{(\lambda^*,\, s)}).
\label{comp_lambda_s}
\end{align}
Moreover, after recognizing that $\hat g_j^{(h)}= \lambda_h \hat g_j^{(s=h)}$ with $\lambda_h = \sin(h)/h < 1$, we have:

\begin{align}
\Delta(\hat g_j^{(h)}) &= \Delta(\lambda_h \hat g_j^{(s=h)})=
 \Delta(\hat g_j^{(\lambda=\lambda_h,\, s)} ) \ge  \Delta(\hat g_j^{(\lambda^*,\, s)}).
\label{comp_lambda_h}
\end{align}
Therefore, the scaled parameter-shift estimator $\hat g_j^{(\lambda^*,\, s)}$ is characterized by a smaller mean 
squared error when compared to both the finite-difference estimator $\hat g_j^{(h)}$  and the (non-scaled) 
parameter-shift estimator $\hat g_j^{(s)}$.

We stress that this result can be straightforwardly generalized also to high-order derivative estimators 
beyond the gradient, by simply replacing the constant $\lambda_h = \sin(h)/h$ which appears in 
Eq.~\eqref{comp_lambda_h} with $\lambda_h = (\sin(h)/h)^d$, where $d$ is the order of the derivative tensor. 
Moreover, we also highlight that this result is independent on the validity of Assumption 1. However, 
if Assumption 1 holds, we can deduce the additional fact that the scaled parameter-shift estimator 
with $s=\pi/2$ is always the optimal choice since it minimizes the mean squared error.

One may wonder why a scaled version of the finite-difference estimator was not considered. The reason is that it would be exactly equivalent to the scaled parameter-shift estimator up to a re-normalization of the scale factor $\lambda$. The choice of scaling the parameter-shift estimator is however more natural since in this case we have that, for a large number of shots $N$, the optimal scaling $\lambda^*$ tends to the trivial limit of $1$.

As a final comment in this section, we note that the forward-difference estimator was not considered in the above comparison.
On the one hand, the forward-difference estimator has a bias that scales linearly with the step size $h$,
as opposed to $h^2$ for the central-difference estimator considered here.
On the other hand, the forward difference estimator can be calculated with roughly half the number
of shifted expectation values. The forward-difference estimator may hence be competitive for particular situations,
depending on the curvature of the cost function and on the choice of shot number and step size.

\section{First- and second-order optimization}\label{Sec:Opt}
\label{sec:optimizers}

In many variational algorithms and in quantum machine learning problems, a loss function (which may depend 
nonlinearly on one or more expectation values) is minimized with respect to the circuit parameters $\bm \theta$. 
For simplicity, in this section we assume that our goal is to minimize a single expectation value $f(\bm \theta)$, 
associated to an observable $M$ as described in  Eq.~\eqref{expval}. More general scenarios can be treated in a 
similar way, with only some classical overhead due to the application of the chain rule for evaluating derivatives 
of composite functions.

\subsection{Gradient descent optimizer}

One of the most common optimizers is {\it gradient descent} (GD). 
According to this algorithm, the parameters are first initialized with an initial (usually random) guess $\bm \theta^{(0)}$
and then, a sequence of updated configurations $\bm \theta^{(1)},\bm \theta^{(2)},\dots, {\bm \theta}^{(T)}$ 
is recursively generated, hopefully converging to a minimum of $f(\bm \theta)$. At each iteration, the GD update rule is:
\begin{align}\label{gd_update}
\bm \theta^{(t)} = \bm \theta^{(t-1)} - \eta \nabla f(\bm \theta^{(t-1)}),
\end{align}
where $\eta>0$ is a real hyper-parameter usually called the {\it learning rate}. Note that this is essentially the same algorithm in 
both the classical and quantum settings,  so the only part of the problem which is intrinsically ``quantum" is the estimation of the gradient.
Indeed, GD  is the default optimizer used in 
most quantum machine learning software frameworks \cite{bergholm2018pennylane, broughton2020tensorflow, qulacs, tequila, luo2019yao}. 
In the previous sections of this work, we have studied different methods 
of estimating the quantum gradient, including the parameter-shift
rule \cite{schuld2019evaluating, mitarai2018quantum, li2017hybrid, mitarai2019methodology}.  

\subsection{Newton optimizer}

 The generalization of the parameter-shift rule to higher-order derivatives that we introduced in Eq.~\eqref{d_rule} 
 and, in particular the analytic evaluation of the Hessian and the Fubini-study metric tensor that we have discussed 
 in the first part of this work, can be directly exploited to implement several second-order optimizers.
Such methods are similar to GD but in place of the first-order update rule of Eq.~\eqref{gd_update} they 
have more advanced iteration rules, which take into account also some information about the curvature of the 
objective function (or of the quantum state) with respect to the variational parameters $\bm \theta$.

For example, the update rule of the Newton optimizer (see, e.g., \cite{gill2019practical}) is:
\begin{align}\label{newton_update}
\bm \theta^{(t)} = \bm \theta^{(t-1)} - 
\eta  [\text{Hess} f(\bm \theta^{(t-1)}) ]^{-1} \nabla f(\bm \theta^{(t-1)}),
\end{align}
where $[\text{Hess} f(\bm \theta) ]^{-1}$ is the inverse of the Hessian matrix, which can be estimated using 
the associated parameter-shift rule presented in Eq.~\eqref{hess_ps_exact_s}. Since the Hessian is an $m \times m$ matrix, 
at each iteration, the number of expectation values which need to be estimated scales as $m^2$.

\subsection{Diagonal Newton optimizer}

 The computational cost of this method can be significantly reduced if one approximates the full Hessian matrix with 
 its diagonal part, such that the update rule for the $j_{\rm th}$  element of $\bm \theta^{(t)}$, simplifies to:
\begin{align}\label{diag_newton_update}
\bm \theta_j^{(t)} = \bm \theta_j^{(t-1)} - 
\eta   g_{j,j}^{-1}(\bm \theta^{(t-1)}) g_j(\bm \theta^{(t-1)}),
\end{align}
where we used notation of the derivative tensor of Eq.~\eqref{eq:g_tensor}. 

The computational cost of this optimizer, which was originally introduced in classical machine learning \cite{becker1988improving}, has a linear scaling with
respect to $m$ and so it can be a competitive alternative to GD. Moreover, the computational overhead of this 
method with respect to the analytic GD optimizer is negligible. Indeed, according to Eq.~\eqref{hess_diag_half}, 
the diagonal of the Hessian can be evaluated with the same $\pm \pi/2$ shifts which would be measured anyway to estimate for the gradient.

\subsection{Quantum natural gradient optimizer}

A different second-order optimizer can be obtained from the concept of quantum natural gradient \cite{stokes2020quantum}. 
The corresponding update rule is very similar to Newton's method, but here the Hessian is replaced by the Fubini-study 
metric tensor defined in Eq.~\eqref{metric-tensor}:
\begin{align}\label{natural_gradient_update}
\bm \theta^{(t)} = \bm \theta^{(t-1)} - 
\eta  [F (\bm \theta^{(t-1)}) ]^{-1} \nabla f(\bm \theta^{(t-1)}).
\end{align}
As we have shown in Section \ref{sec:metric-tensor}, all the elements of metric tensor $F_{i,j}$ can be 
analytically estimated with the parameter-shift rule derived in Eq.~\eqref{metric_rule}. 

\subsection{Regularization of second-order methods}
\label{sec:regularization}
All the previous second-order methods share the same structure of the update rule 
$\bm \theta^{(t)} = \bm \theta^{(t-1)} - \eta  A^{-1} \nabla f(\bm \theta^{(t-1)})$, where $A$ can be the Hessian, 
the metric tensor or some diagonal approximation. In the field of classical optimization, 
it is well known  that, if $A$ is not positive-definite, the optimizer can converge to a maximum instead of a minimum 
\cite{gill2019practical}. Moreover, if one or more eigenvalues are close to zero, the norm of the inverse matrix 
can be so large that the algorithm becomes unstable. Therefore it is 
a common practice to regularize $A$ in such a way to make it sufficiently positive. 
For example one can replace $A$ with $A'= A + \epsilon \mathbbm{1}$, for some positive parameter $\epsilon>0$, 
which corresponds to shifting all the eigenvalues by $+\epsilon$.
Alternatively, one can regularize each eigenvalue $\lambda_k$ of $A$ by replacing it with $\lambda_k'= \max(\lambda_k, \epsilon)$, without changing the associated eigenvector. In our experiments we used the previous method, however, from additional numerical simulations (see Appendix~\ref{Sec:NumericsExtra}, Sec.~\ref{sec:different_regularization}) we noticed that the second method can be
much more efficient since it only perturbs $A$ in the necessary subspace and keeps $A$ invariant whenever it is sufficiently positive.

Finally we mention the approach which was recently theoretically proposed in Ref.~\cite{huembeli2020characterizing}, where the inverse of the maximum eigenvalue of the Hessian is used as a learning rate. In the notation of our work, this approach could be interpreted as a particular regularization method in which $A$ is replaced by $A'=\max(\lambda_k) \mathbbm{1}$. It is worth to underline that  Ref.~\cite{huembeli2020characterizing} contains a comprehensive study on the spectrum of the Hessian and about its application for analyzing the landscape of the cost function. This general analysis is beyond the aim of our work, which is instead more focused on the experimental estimation of the Hessian and its usage in the Newton optimizer.

We also highlight the detailed analysis of Ref.~\cite{van2020measurement}
in which the error propagation associated to the matrix inversion of $A'$
(the regularized Hessian or metric tensor) is studied in the context of
second-order optimization methods. Quite interestingly, according to Ref.~\cite{van2020measurement}, second-order optimizers have a relatively low sensitivity 
to the errors in the matrix elements of $A'$ and so they can still converge towards a minimum even when the Hessian or the metric tensor are statistically noisy. This fact is consistent with the experimental results reported Sec.~\ref{Sec:Numerics}.

\section{Numerical and hardware experiments}\label{Sec:Numerics}

This section contains details of experiments run on simulators and hardware. 
We use the PennyLane software
library~\cite{bergholm2018pennylane} to construct a simple
variational quantum circuit and access its expectation value and gradient. The
circuit is evaluated using PennyLane's built-in simulator as well as on hardware
by using integration with the IBM Quantum Experience to execute on the 
\texttt{ibmq\_valencia} and \texttt{ibmq\_burlington} five-qubit chips.

%\begin{figure}[t!]
%  \centering
%  \mbox{
%  \Qcircuit @C=1em @R=.7em {
%& \gate{R_{X}(\theta_{1})} & \ctrl{1} & \qw & \qw & \qw & \qw \\
%& \gate{R_{X}(\theta_{2})} & \targ  & \targ & \targ & \qw & \measuretab{\hat{\sigma}_{z}} \\
%& \gate{R_{X}(\theta_{3})} & \qw  & \ctrl{-1} & \qw & \qw & \qw \\
%& \gate{R_{X}(\theta_{4})} & \qw  & \qw & \ctrl{-2} & \targ & \qw \\
%& \gate{R_{X}(\theta_{5})} & \qw  & \qw & \qw & \ctrl{-1} & \qw \\
%  }}
%  \caption{The variational circuit run on simulator and hardware to investigate the
%  findings in this paper.  
%  }\label{fig:circuit}
%\end{figure}
\begin{figure}[t!]
  \includegraphics[scale=0.2]{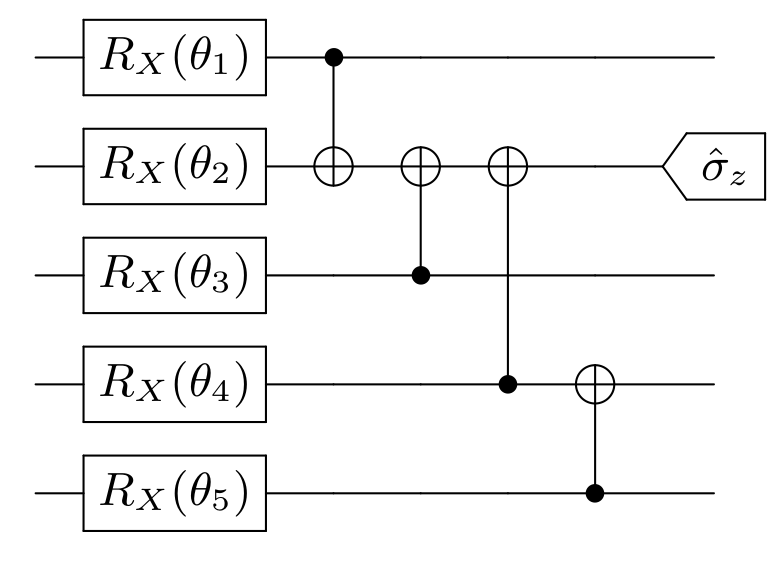}
  \caption{The variational circuit run on simulator and hardware to investigate the
  findings in this paper.  
  }\label{fig:circuit}
\end{figure}

The circuit has five parameters
$\bm{\theta} = (\theta_{1}, \theta_{2}, \theta_{3}, \theta_{4}, \theta_{5})$,
as shown in Fig.~\ref{fig:circuit}. It consists of a collection of single
qubit Pauli-$X$ rotations, an entangling block, and a measurement of the Pauli-$Z$
observable on the second qubit. The circuit is chosen to provide a non-trivial
gradient with a low depth and the entangling block is set to match the topology
of the hardware devices. We use a fixed set of parameters
$\bm{\theta}$, which is detailed in Appendix~\ref{Sec:NumericsExtra} along with
the resultant analytic expectation value and gradient.

\subsection{Estimating the gradient}

\begin{figure}[t!]
  \centering
  \includegraphics[scale=0.53]{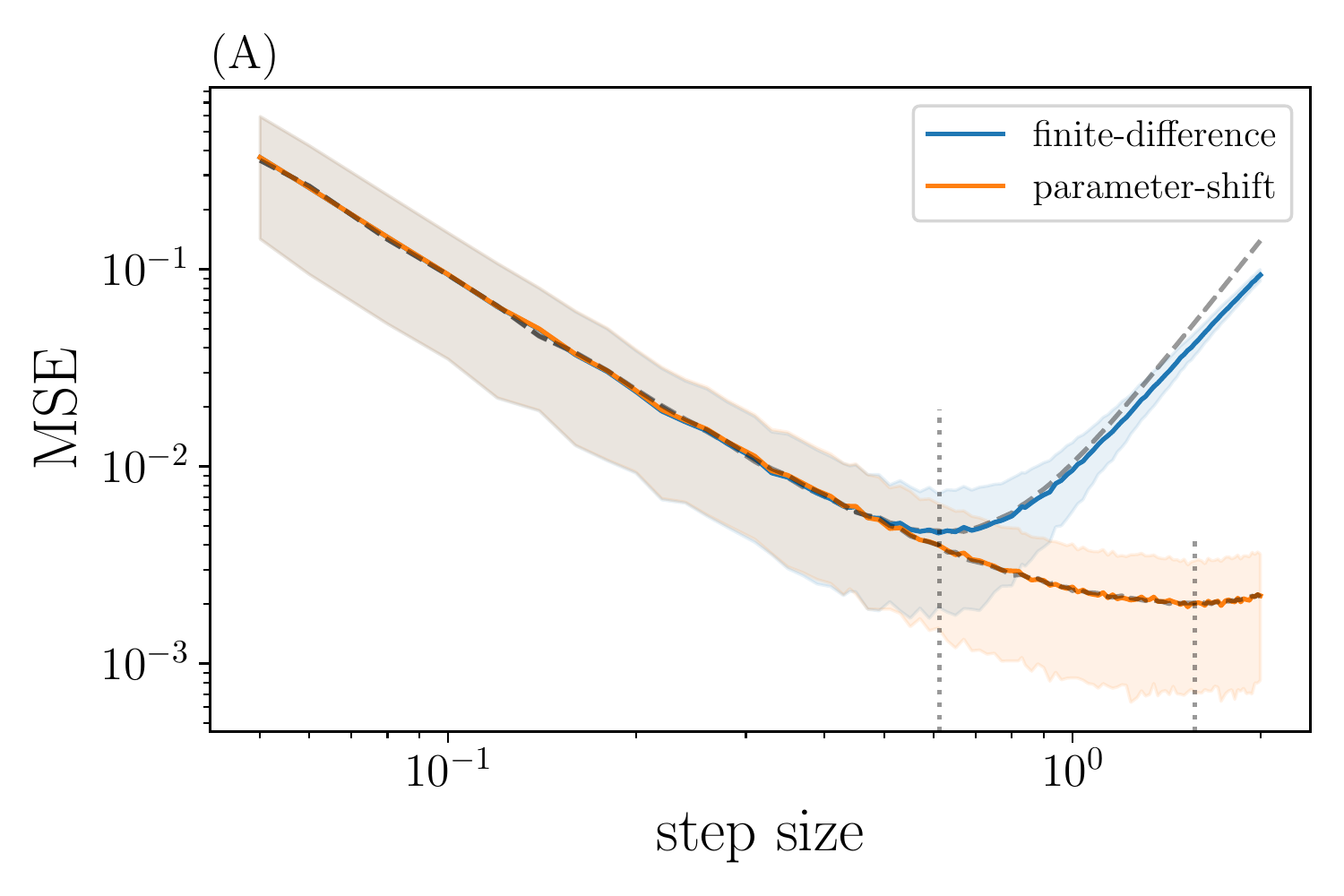}
  \includegraphics[scale=0.53]{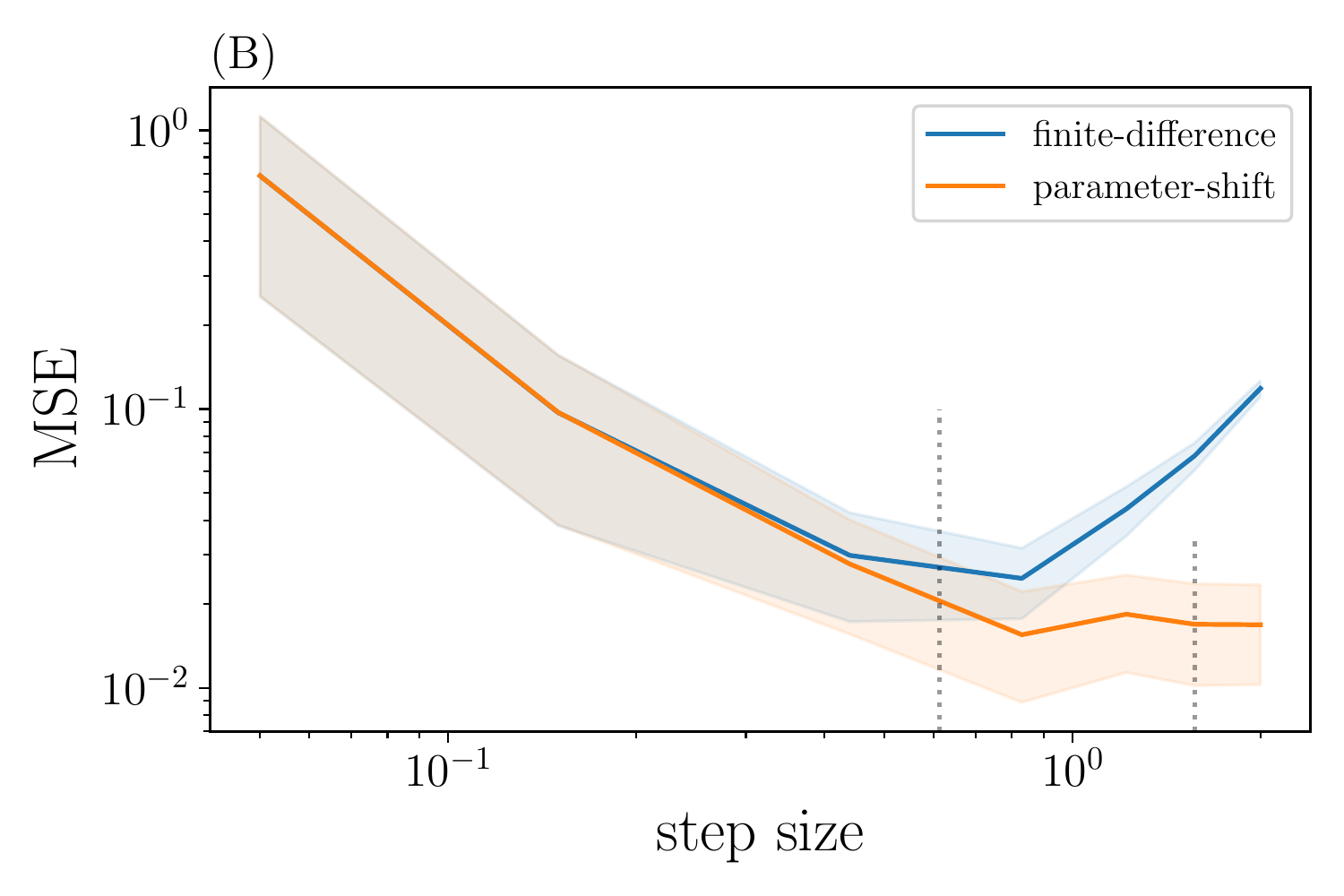}
  \caption{Mean squared error (MSE) for different choices of step size when
  estimating the gradient using the finite-difference and parameter-shift estimators
  with $10^{3}$ shots used to evaluate expectation values.
  The solid lines show the MSE while the shaded regions illustrate the range
  of values within one standard deviation. Plot (A) compares the estimators when run
  on a simulator, while Plot (B) compares the estimators when run on
  \texttt{ibmq\_valencia}.
  Additional gray lines are calculated from
  theoretical predictions: the dashed lines show the expected behavior of the MSE and the dotted vertical
  lines show the expected optimal choice of step size: $h^{*} = 0.613$ and
  $s^{*} = \pi / 2$.
  }\label{fig:h_and_err_vs_shots}
\end{figure}

We now investigate the performance of gradient estimators in the finite-shot setting,
as discussed in Sec.~\ref{Sec:FiniteShots}. Figure~\ref{fig:h_and_err_vs_shots} (A)
shows how the MSE $\Delta(\hat{\bm{g}})$ scales
with step size for the finite-difference and parameter-shift estimators when
expectation values are estimated on a simulator with $10^{3}$ shots. The MSE is
calculated over $10^{3}$ repetitions and seen to coincide closely with the
theoretical predictions provided earlier. Note that the finite-difference curve
starts to diverge from the predicted behavior for large step sizes due to the
Taylor-series approximation breaking down in Eq.~\eqref{bias_h}.

It is hence clear from the figure that the
parameter-shift method with $s = \pi / 2$ is the best performing estimator in this
setting. We investigate the relative performance of the two gradient estimators
further using simulations in Appendix~\ref{Sec:NumericsExtra}, where we see that the parameter-shift
method has the lower MSE for $N > 50$ shots. For low shot numbers, the
scaled parameter-shift method mentioned in Sec.~\ref{Sec:Rescaled} can also be
used.

\begin{figure}[t!]
  \centering
  \includegraphics[scale=0.6]{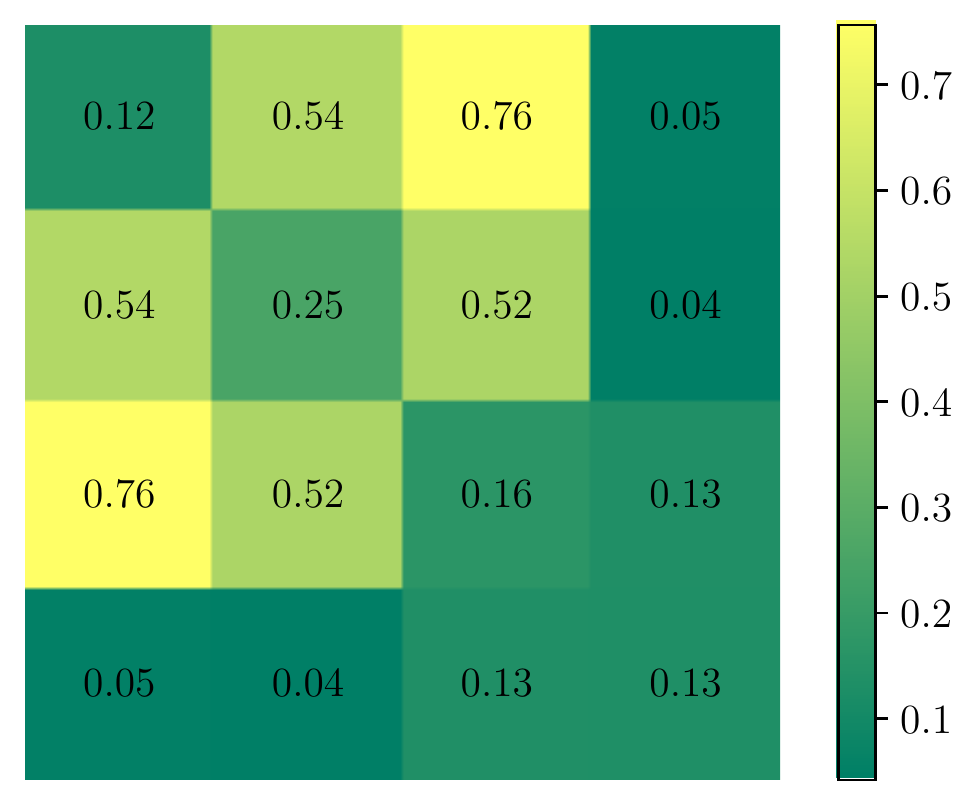}
  \caption{Relative error for using the parameter-shift estimator to approximate
  the Hessian matrix on the \texttt{ibmq\_valencia} hardware device.
  Only the upper left $4 \times 4$ block is considered, since the exact Hessian is zero
  on the final row and column.
  }\label{fig:hessian_hw}
\end{figure}

\begin{figure*}[t!]
  \centering
  \includegraphics[scale=0.43]{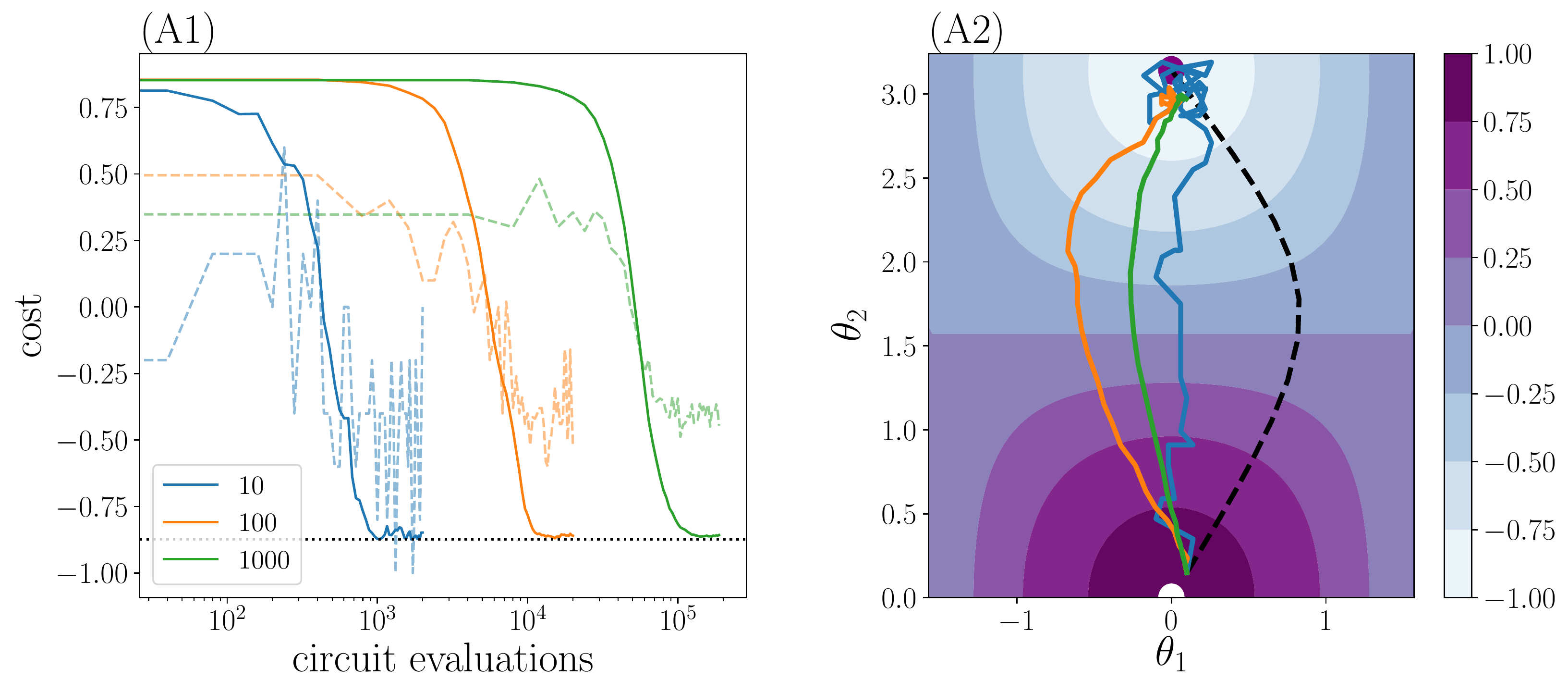}
  \includegraphics[scale=0.43]{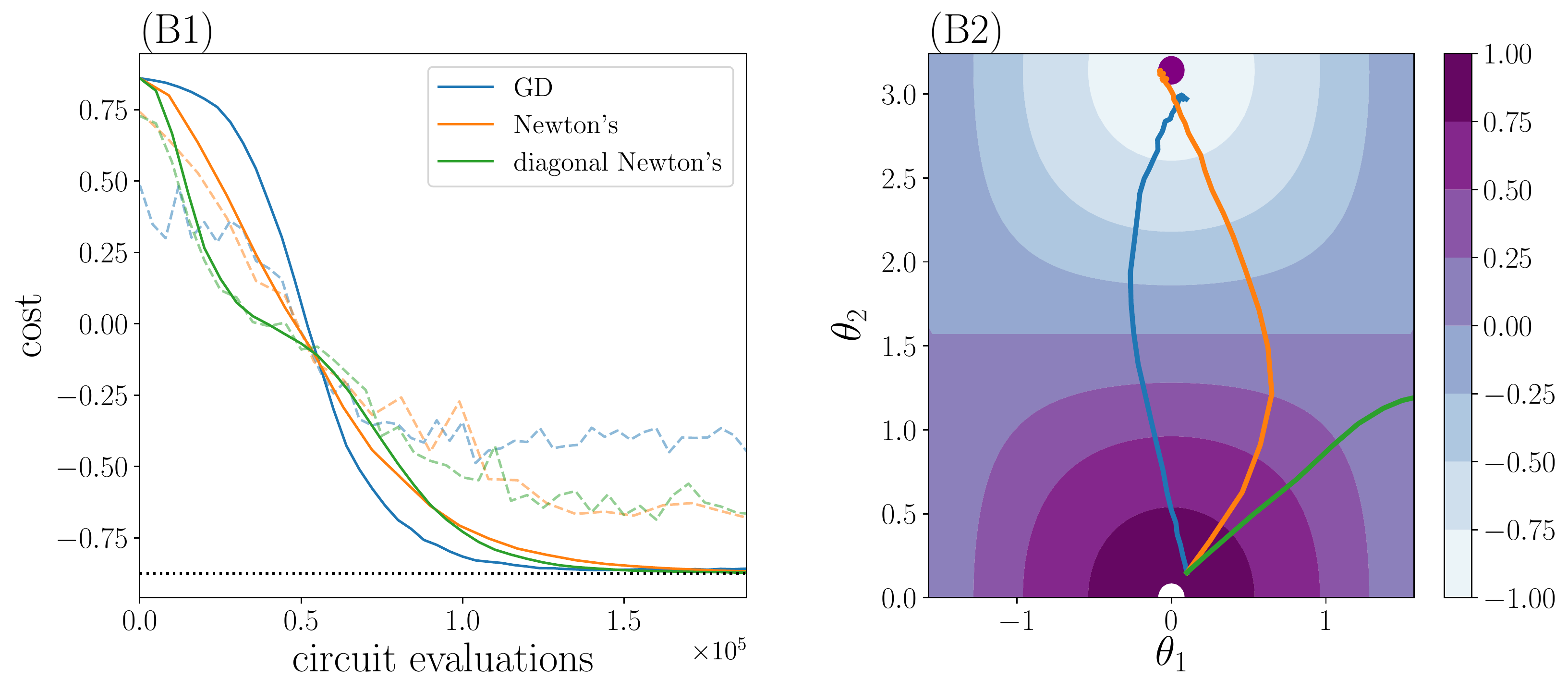}
  \caption{A comparison of optimizers with circuit evaluation performed on the
  \texttt{ibmq\_burlington} and \texttt{ibmq\_valencia} hardware devices. 
  The first row of plots (experiment A) focuses on the gradient descent (GD) optimizer for a varying
  number of shots $N$ per expectation value measurement, while the second row of
  plots (experiment B) compares different optimizers discussed in Sec.~\ref{Sec:Opt} when using
  $N=1000$. Plots (A1) and (B1) show the cost function $f(\bm{\theta})$
  in terms of the total number of circuit evaluations on the device and
  Plots (A2) and (B2) show the path taken by the optimizer through the
  $\theta_{1}$-$\theta_{2}$ space.
  In (A1) and (B1), the dashed lines give the cost function evaluated on hardware
  and the solid
  lines give the cost function evaluated on an exact simulator with the same
  $\bm{\theta}$.
  The dotted line gives the analytic minimum of $-0.874$.
  In (A2) and (B2), the contrasting white and purple circles highlight the maximum and
  minimum, while the dashed line in (A2) is the path taken by the GD optimizer with
  access to exact expectation values.
  }\label{fig:optimizers}
\end{figure*}

Figure~\ref{fig:h_and_err_vs_shots} (B) compares the gradient estimators using
the \texttt{ibmq\_valencia} device with $10^{3}$ shots per expectation value. The
MSE is calculated over $16$ repetitions due to limited availability of the hardware.
We observe a similar qualitative behavior to the simulator-based results in
Fig.~\ref{fig:h_and_err_vs_shots} (A) but with the MSE offset upward by roughly
one order of magnitude, likely due to systematic errors in the device.

\subsection{Estimating the Hessian}

We use the results presented in Sec.~\ref{Sec:Hessian} to measure the Hessian
matrix using the parameter-shift method. The Hessian of the circuit in
Fig.~\ref{fig:circuit} is a $(5 \times 5)$-dimensional symmetric matrix with $10$
off-diagonal terms and $5$ along the diagonal. The parameter-shift method with
shift $s=\pi / 2$ was used to estimate the off-diagonal terms, requiring
measurement of $40$ expectation values. The diagonal terms were estimated with
$s = \pi / 4$ which required measuring $11$ expectation values.
The exact value of the Hessian matrix is provided in
Appendix~\ref{Sec:NumericsExtra}, which also includes a simulator-based
comparison of the MSE for the Hessian when estimated using the finite-difference
and parameter-shift methods.
The exact Hessian matrix consists of a $4 \times 4$ block of non-zero terms
and a final row and column of zeros due to the expectation value being
independent of $\theta_{5}$. We estimated the Hessian matrix on hardware using
\texttt{ibmq\_valencia} with $N=10^{3}$. Figure~\ref{fig:hessian_hw} shows the relative error
for the $4 \times 4$ block, while the maximum squared error for second derivatives
including the $\theta_{5}$ term was $8.12 \times 10^{-4}$.

From Fig.~\ref{fig:hessian_hw}, one can deduce that off-diagonal terms tend to have a larger relative error. Since most of the weight of the exact Hessian is on the diagonal (see Appendix~\ref{Sec:NumericsExtra}) the error on the off-diagonal terms may be due to statistical fluctuations which have a larger impact on small matrix elements. However, given the limited amount of data, it is hard to distinguish between the errors which are caused by the experimental device and those which just correspond to statistical fluctuations.

\subsection{Optimization}

The GD, Newton and diagonal Newton optimizers discussed in Sec.~\ref{Sec:Opt}
are now used to minimize the expectation value $f(\bm{\theta})$ of the circuit
in Fig.~\ref{fig:circuit} when run on hardware using the \texttt{ibmq\_valencia}
and \texttt{ibmq\_burlington} devices. We select $\theta_{1}$ and $\theta_{2}$ as
trainable parameters and leave $\theta_{3}$, $\theta_{4}$, and $\theta_{5}$ fixed, as
well as choosing a constant learning rate of $\eta = 0.4$,
as detailed in Appendix~\ref{Sec:NumericsExtra}.

A comparison of the performance of the GD optimizer when using different shot numbers
$N$ to evaluate expectation values is provided in row A of Fig.~\ref{fig:optimizers}. Plot (A1) shows the cost function $f(\bm{\theta})$ evolving with the total
number of circuit evaluations on the device, while Plot (A2) illustrates the path taken
by the optimizer through the $\theta_{1}$-$\theta_{2}$ space. The starting point of
the optimizer is $(\theta_{1}, \theta_{2}) = (0.1, 0.15)$ and an exact GD optimizer
tends to the value $(\theta_{1}, \theta_{2}) = (0, \pi)$.

These plots highlight some interesting features of optimization on hardware.
We see that the minimum can be approached even with a low shot number of $N=10$,
requiring orders of magnitude fewer circuit evaluations from the device than the
$N=100$ and $N=1000$ cases. However, the $N=10$ case is noisy and oscillates around
the minimum point. This suggests that an adaptive shot number $N$ may be useful
in practice with a low shot number initially and an increase in $N$ as the optimizer
approaches a minimum, as has been discussed in recent
works~\cite{sweke2019stochastic,arrasmith2020operator,kubler2020adaptive}.
It is also important to note that the optimizer is able to approach the expected
minimum even though the cost function available on the device is noisy, indicating
that the direction of the gradient is still an accessible signal. Nevertheless, the
gradient direction is clearly still prone to error: note that in Fig.~\ref{fig:optimizers}
the hardware-based paths are different from the ideal simulated path.

Row B of Fig.~\ref{fig:optimizers} shows the result of using
the GD, Newton
and diagonal Newton optimizers on hardware with $N=10^{3}$ and with the same
starting point of $(\theta_{1}, \theta_{2}) = (0.1, 0.15)$. The Newton optimizer
requires evaluation of a $(2 \times 2)$-dimensional Hessian at each iteration step
which is realised with $9$ expectation value measurements, while the diagonal Newton 
optimizer requires $5$ expectation value measurements per iteration step. This
contrasts with the GD optimizer which requires $4$ evaluations per step. These
differences are factored into Plot (B1) of Fig.~\ref{fig:optimizers}, which plots
the cost function $f(\bm{\theta})$ in terms of the total number of circuit
evaluations taken on the
device.

We observe a comparable performance between each of the optimizers, but it
is important to note that the learning rate hyperparameter and the regularization 
method can change the performance
of each optimizer.
One may also wonder why, for a large number of circuit evaluations, GD seems to be equivalent
or even better than second-order methods. The supplementary theoretical analysis reported in  Appendix~\ref{Sec:NumericsExtra} suggests that this is strongly related to the choice of the regularization method (see in particular Sec.~\ref{sec:different_regularization}).
The dramatic dependence of Newton optimizers on the regularization strategy is a drawback which is well known
in the field of classical optimization \cite{gill2019practical}. This is a practical issue which must be taken into account also in the quantum setting.

The paths taken by each optimizer are drawn in Plot (B2) of
Fig.~\ref{fig:optimizers}. This highlights the key qualitative difference between
the optimizers, which take paths based on the different information they have
available, such as the local curvature information for the optimizers using the Hessian.
Note that the diagonal Newton optimizer tended to the $(\pi, 0)$ minimum when
evaluated on hardware. This behaviour was due to the presence of noise in the device:
all optimizers tend to the $(0, \pi)$ minimum with access to an ideal simulator, as
shown in Appendix~\ref{Sec:NumericsExtra}.

The main scope of our experiments was to show the  practical feasibility of GD in the regime of high statistical noise and to give a proof-of-principle demonstration of second-order methods on real hardware. Consistently with this scope, we used a very simple circuit but one should be aware that, in real applications, the circuits are usually quite different (larger and more structured). 
A quantitative and detailed benchmarking of the different optimizers is left for a future work. On this subject, we mention the recent theoretical analysis reported in  Ref.~\cite{wierichs2020avoiding} and  Ref.~\cite{huembeli2020characterizing}.
\\

\section{Conclusions}

We have derived a generalization of the parameter-shift rule for evaluating derivatives of arbitrary
order. We have studied how such derivatives are affected by the statistical noise which is due to a finite 
number of experimental measurements, testing our predictions with numerical simulations and real quantum experiments. 
We have also experimentally  tested how the generalized parameter-shift rules can be used to efficiently implement second-order optimization methods on near-term quantum computers.

The results presented in this work could pave the way to interesting future research directions. For example,
the strong similarity between the classical {\it round-off} error the quantum statistical error, which naturally emerged in our
analysis of derivative estimators, is probably worth being further explored and generalized. Moreover, the analytic second-order optimizers, which in this work we tested with simple examples, could be subject to a more systematic benchmarking analysis in order to understand
their competitive potential with respect to first-order methods. 

\acknowledgments 

We thank Josh Izaac for his help with numerical and hardware experiments. We also thank Maria Schuld for useful comments.
This material is based upon work supported by the Defense Advanced Research Projects Agency (DARPA) under Agreement No. HR00112090015 (Backpropagating Through Quantum Computers).

% References

%\nocite{*}
% whether or not they are actually referenced in the text. 

%\bibliographystyle{apsrev4-1}
%\bibliographystyle{abbrv} 
\bibliographystyle{unsrt}

\bibliography{references}

\newpage 
\appendix

\section{Numerics and hardware experiments: further details}\label{Sec:NumericsExtra}

\subsection{Problem setting}

We use a fixed set of parameters generated from
the uniform distribution over the interval $[0, 2 \pi]$,
\begin{equation}
\bm{\theta} = (2.739, 0.163, 3.454 , 2.735, 2.641)
\end{equation}
Using an exact simulator, the expectation value, its gradient, and its Hessian
can be evaluated as
\begin{align}
f(\bm{\theta}) &= -0.794,  \\ 
\bm{g}(\bm{\theta}) &= (-0.338,  0.130,  0.256, -0.342,  0 ),  \\
H(\bm{\theta}) &= \left( \begin{array}{ccccc}
 0.794 &  0.055 &  0.109 & -0.145 &  0\\
 0.055 &  0.794 & -0.042 &  0.056 & 0\\
 0.109 & -0.042 &  0.794 &  0.110  &  0\\
-0.145 &  0.056 &  0.110  &  0.794 &  0\\
 0   & 0   &  0   &  0   & 0\\
\end{array}
\right)
\end{align}
Note that the final element of the gradient vector is exactly zero. Simulations
were carried out in this paper using the \texttt{default.qubit} device in PennyLane
with a finite number of shots. Hardware evaluation was performed using the
\texttt{qiskit.ibmq} device in the PennyLane-Qiskit plugin.

\begin{figure}[t!]
  \centering
  \includegraphics[scale=0.5]{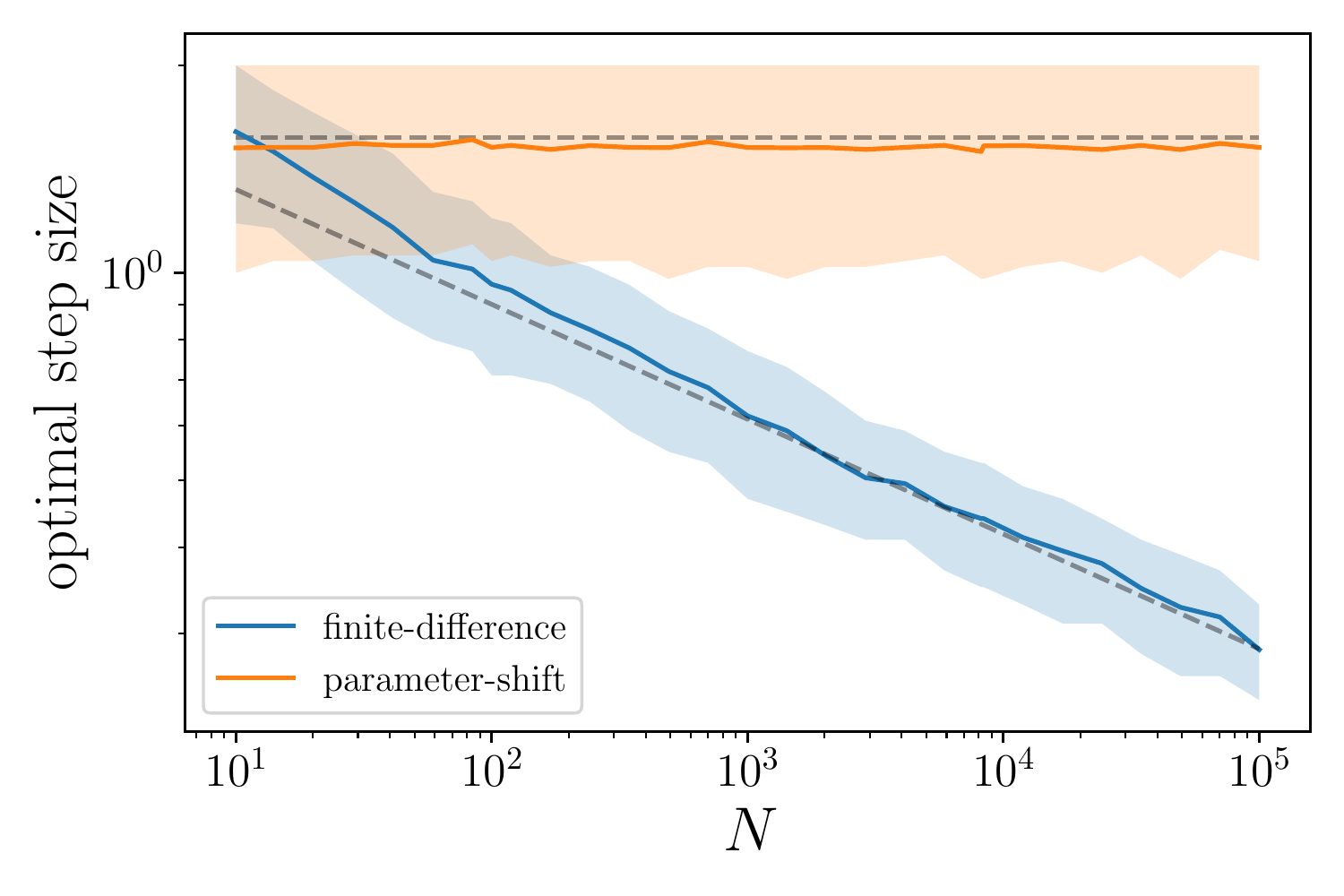}
  \caption{Optimal step size for the finite-difference and parameter-shift
  gradient estimators for a range of shot numbers $N$. The shaded regions display
  the range of step sizes resulting in an MSE that is within $20 \%$ of the
  minimum observed value, while the solid lines are the medians of these
  regions. The dashed gray lines are the predicted values for the optimal step size.
  }\label{fig:h_and_err_vs_shots-h}
    \includegraphics[scale=0.5]{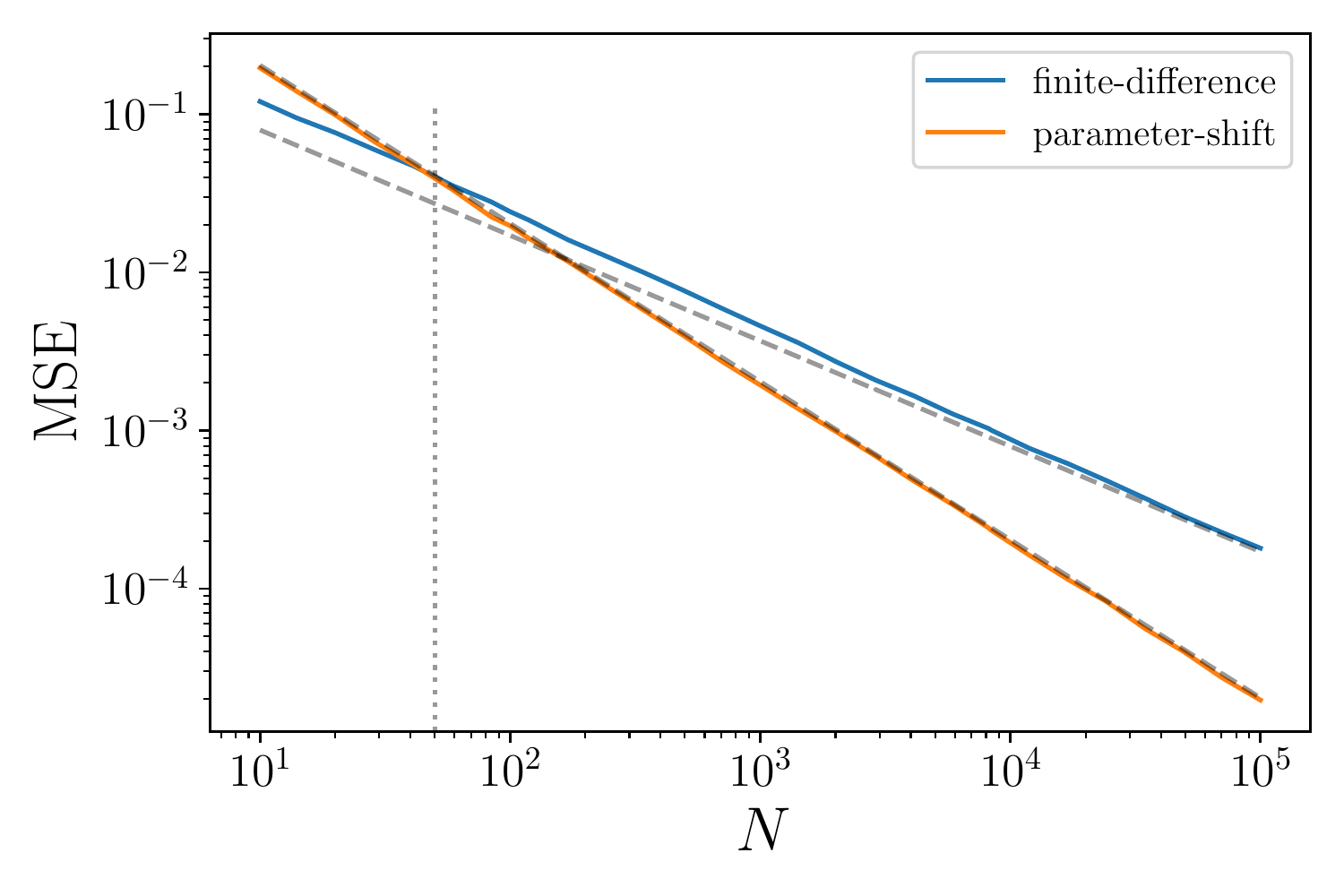}
  \caption{Mean squared error (MSE) for different choices of shot number $N$ when
  using the finite-difference and parameter-shift estimators
  with corresponding optimal choices of step size to estimate the gradient. The
  solid lines are shown from simulation while the dashed gray lines are
  the predicted scaling. The dotted vertical line indicates the crossover-point
  between the two methods at $N \approx 50$ shots.
  }\label{fig:h_and_err_vs_shots-N}
\end{figure}

\subsection{Estimating the gradient}

Figure~\ref{fig:h_and_err_vs_shots-h} shows the optimal step size for both the
finite-difference and parameter-shift gradient estimators over a range of
shot numbers $N$ used to measure expectation values on simulator. In both cases, 
the gradient was estimated over a fixed set of candidate choices for the step size
and the one with the smallest MSE was selected. The plots also contain dashed lines
illustrating the predicted value of the optimal step size. For small $N$, the optimal
step size for the finite-difference estimator begins to deviate from the prediction.
The reason for this deviation is that, for small $N$, the optimal step size $h^*$ is
so large that the Taylor expansion around $h\simeq0$ is not valid anymore. Figure~\ref{fig:h_and_err_vs_shots-N} shows the MSE for the gradient estimators
when using their optimal step sizes over a range of shot numbers $N$. 
In both Figs.~\ref{fig:h_and_err_vs_shots-h} and~\ref{fig:h_and_err_vs_shots-N},
the MSE is calculated using $1000$ repetitions.

% theoretical predictions are calculated using Eqs.~\eqref{bias_h},~\eqref{intermediate_var_h}, and~\eqref{h_opt} 

\subsection{Estimating the Hessian}

\begin{figure}[t!]
  \centering
  \includegraphics[scale=0.5]{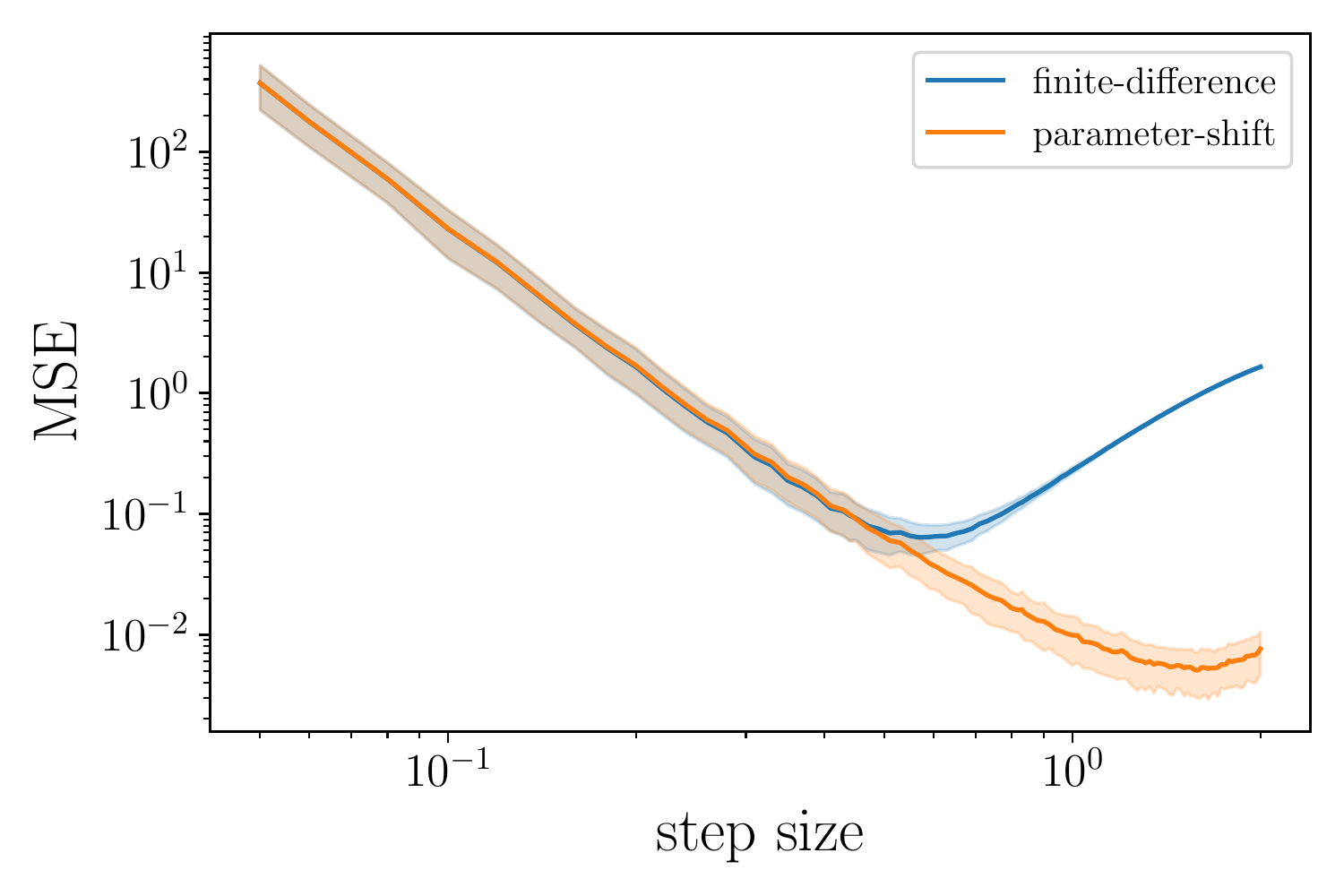}
  \caption{Mean squared error (MSE) for different choices of step size when
  estimating the Hessian using the finite-difference and parameter-shift estimators
  with $10^{3}$ shots used to evaluate expectation values on a simulator.
  The solid lines show the
  MSE while the shaded regions illustrate the range of values within one standard
  deviation.
  }\label{fig:fd_vs_ps_hess}
  \includegraphics[scale=0.5]{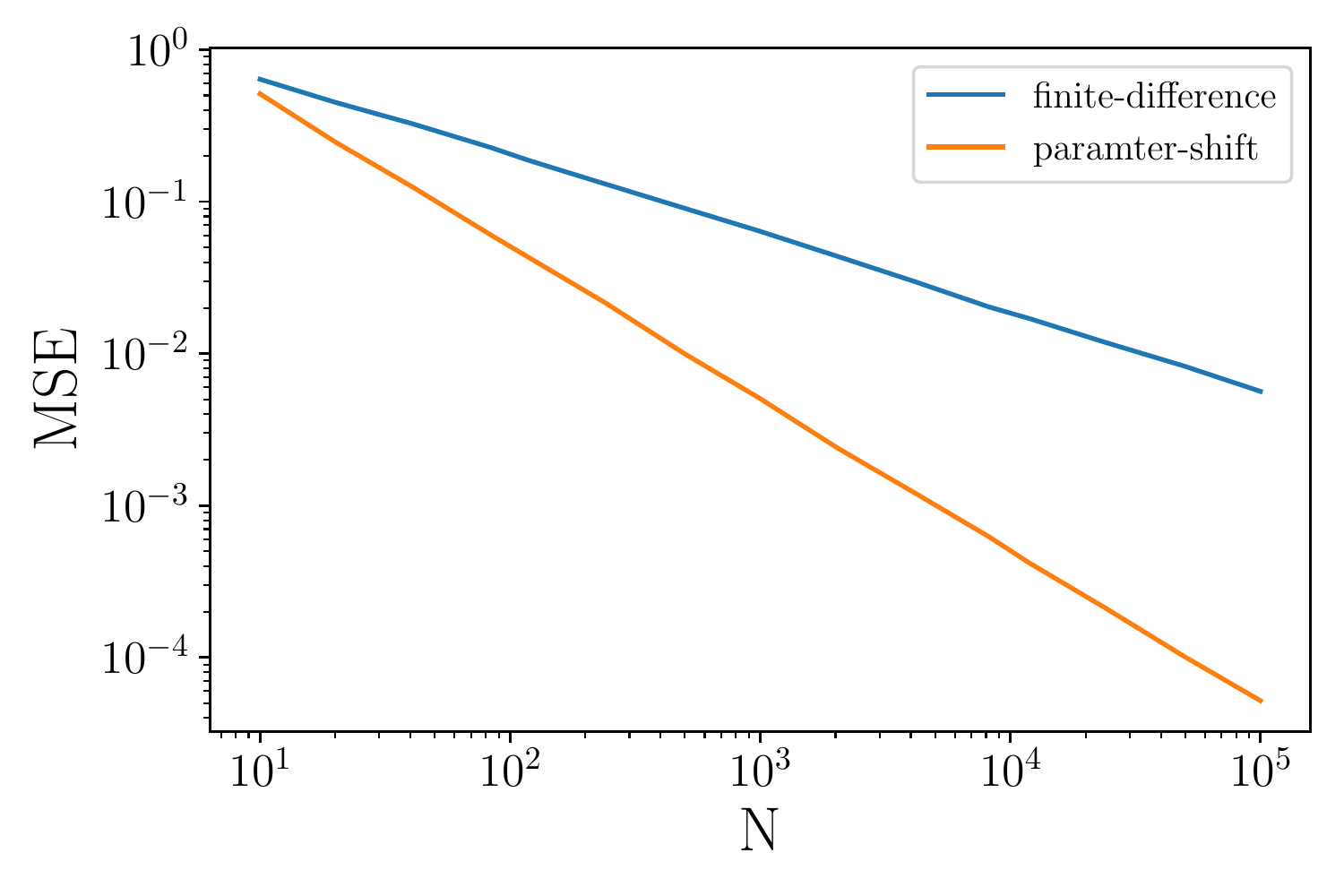}
  \caption{Mean squared error (MSE) for different choices of shot number $N$ when
  using the finite-difference and parameter-shift estimators
  with optimal choices of step size to estimate the Hessian. A power-law fit to each
  line gives a scaling of $N^{-0.5013}$ and $N^{-1.0008}$ for the finite-difference
  and parameter-shift methods, respectively.
  }\label{fig:fd_vs_ps_N}
\end{figure}

Figure~\ref{fig:fd_vs_ps_hess} illustrates the MSE when the Hessian is estimated
using the finite-difference and parameter-shift methods for varying step sizes
when expectation values are
estimated on a simulator with $10^{3}$ shots and with $10^{3}$ repetitions to
calculate the average. The parameter-shift estimator is calculated using
Eq.~\ref{hess_ps_exact_s} and the finite-difference estimator of the Hessian is
taken to be \cite{abramowitz1972abramowitz}
\begin{align}
\hat{g}_{j_1, j_2}^{(h)}(\bm \theta)  
=& [f(\bm \theta + h (\e_{j_1} + \e_{j_2})) - f(\bm \theta + h (-\e_{j_1} + \e_{j_2}))  \nonumber  \\
& -f(\bm \theta + h (\e_{j_1} - \e_{j_2}))+ f(\bm \theta - h (\e_{j_1} + \e_{j_2}))] \nonumber \\
& /4 h^{2}.
\end{align}
Figure~\ref{fig:fd_vs_ps_N} compares both estimators of the Hessian for varying
shot numbers $N$ when each estimator uses the corresponding optimal step size.

\subsection{Second-order optimization}

We minimize the expectation value $f(\bm{\theta})$ with
$
\bm{\theta} = (\theta_{1}, \theta_{2}, 3.454, 2.735, 2.641),
$
where $\theta_{1}$ and $\theta_{2}$ are trainable parameters. The analytic minimum
is equal to
\begin{equation}
\min_{\theta_{1}, \theta_{2}} f(\bm{\theta}) = -0.874.
\end{equation}
Extremal points occur alternately when $\theta_{1}$ and $\theta_{2}$ are
multiples of $\pi$.

In the results provided, \texttt{ibmq\_burlington} was used
when investigating the GD optimizer and \texttt{ibmq\_valencia} was used when
investigating the Newton and diagonal Newton optimizers. A learning rate of
$\eta = 0.4$ was adopted for all optimizers.

\begin{figure*}[t!]
  \centering
  \includegraphics[scale=0.45]{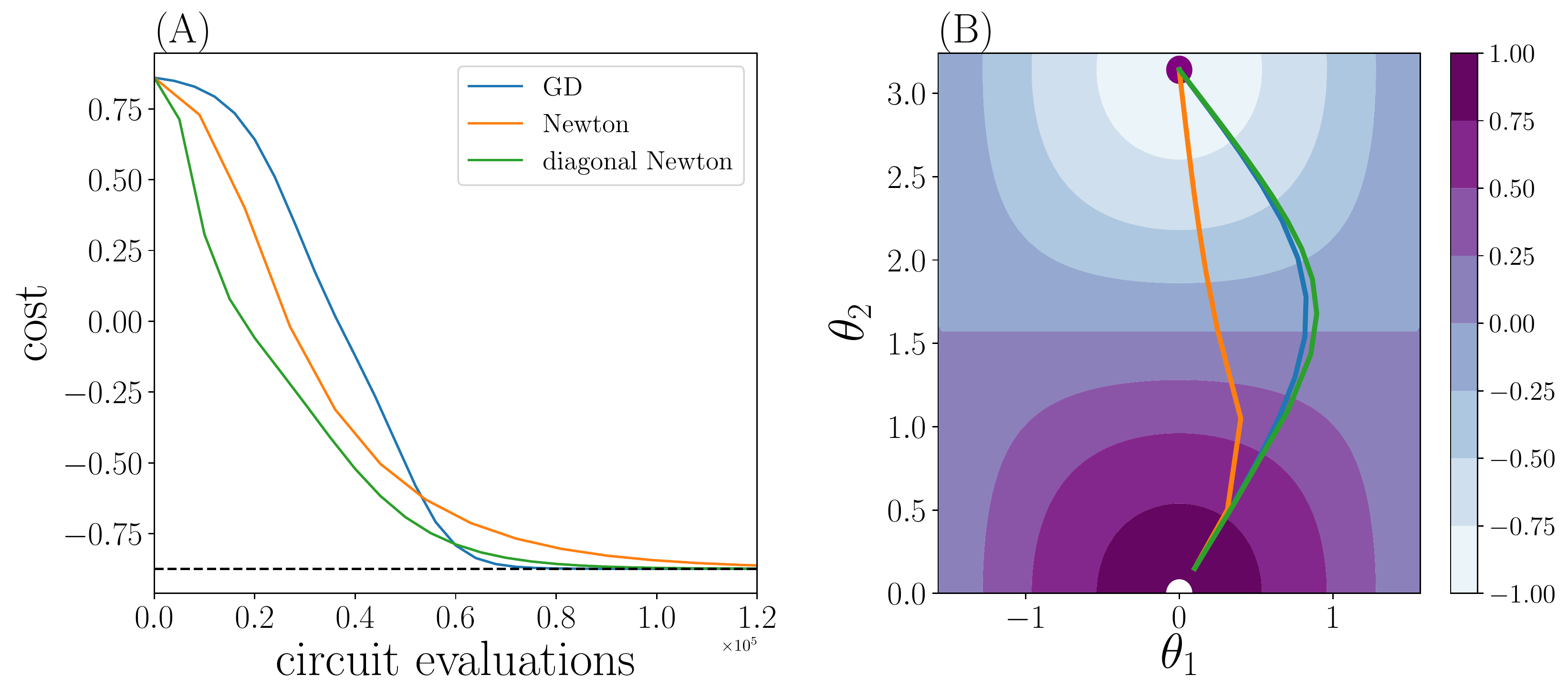}
  \caption{A comparison of different optimizers discussed in Sec.~\ref{Sec:Opt}
  with circuit evaluation performed on PennyLane's noise-free \texttt{default.qubit} simulator.
  Plot (A) shows the cost function $f(\bm{\theta})$
  in terms of the total number of circuit evaluations and
  Plot (B) shows the path taken by the optimizer through the
  $\theta_{1}$-$\theta_{2}$ space.
  In (A), the dotted line gives the analytic minimum of $-0.874$
  and in (B), the contrasting white and purple circles highlight the maximum and
  minimum.
  }\label{fig:optimizers_sim}
\end{figure*}

Figure~\ref{fig:optimizers_sim} compares the GD, Newton and diagonal Newton
optimizers with circuit evaluation on a noise-free simulator.

\subsection{Using a different regularization}
\label{sec:different_regularization}
\begin{figure*}[t!]
  \centering
  \includegraphics[scale=0.45]{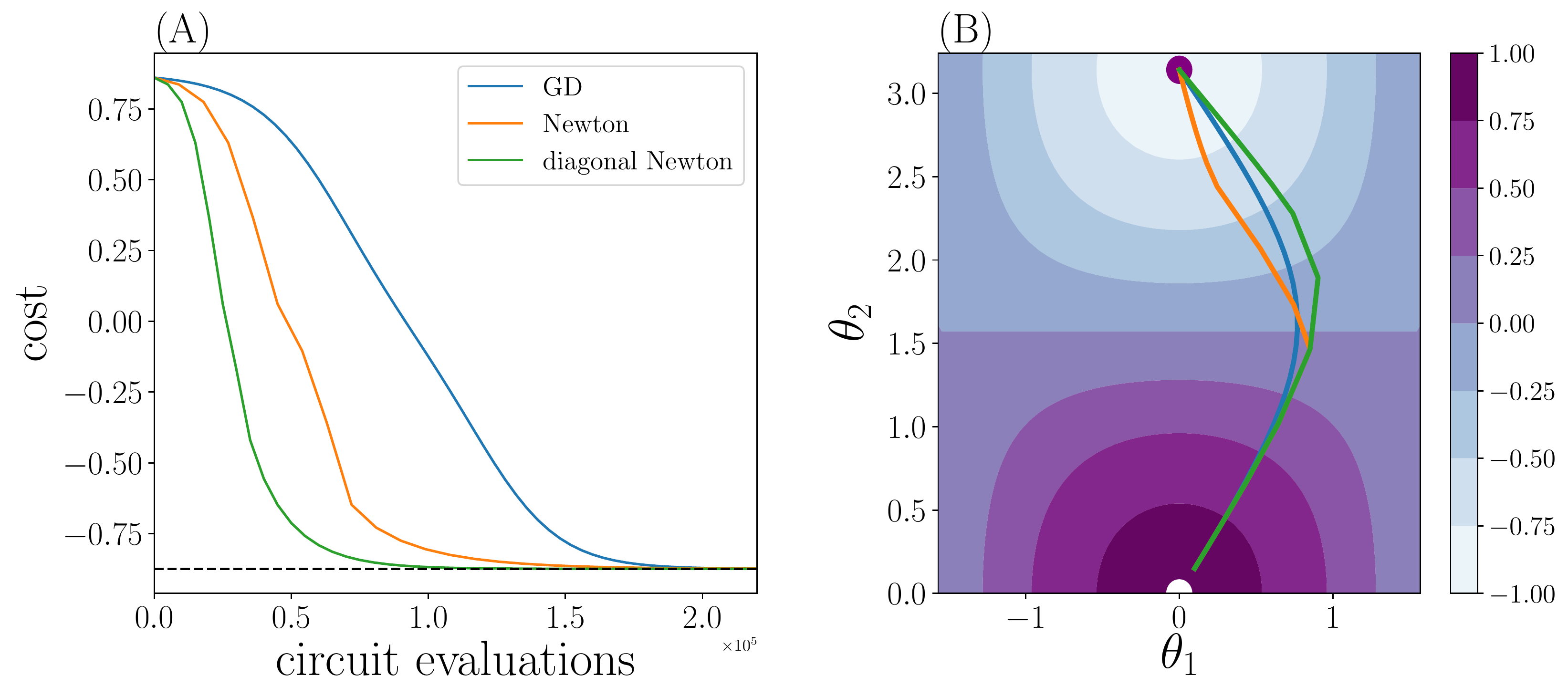}
  \caption{A comparison of different optimizers discussed in Sec.~\ref{Sec:Opt}
  with circuit evaluation performed on PennyLane's noise-free \texttt{default.qubit} simulator.
  Plot (A) shows the cost function $f(\bm{\theta})$
  in terms of the total number of circuit evaluations and
  Plot (B) shows the path taken by the optimizer through the
  $\theta_{1}$-$\theta_{2}$ space.
  In (A), the dotted line gives the analytic minimum of $-0.874$
  and in (B), the contrasting white and purple circles highlight the maximum and
  minimum. With respect to the simulation of Fig.~\ref{fig:optimizers_sim}, in this case a different regularization method is used as discussed in the text.
  }\label{fig:optimizers_sim2}
\end{figure*}

As discussed in Sec.~\ref{sec:regularization}, one can use different regularization methods. In the previous examples we used $\textrm{Hess}(f) \rightarrow \textrm{Hess}(f) + \epsilon \mathbbm{1}$ with $\epsilon>0$ which, in the specific cases of 
Figs.~\ref{fig:optimizers_sim} and \ref{fig:optimizers}, was set to a relatively large value ($\epsilon=1$) to compensate for the large negativity of the initial Hessian matrix.
In Fig.~\ref{fig:optimizers_sim2} instead, we numerically simulate the same optimization problem but we use a different regularization method: we replace each eigenvalue $\lambda_k$ of $\textrm{Hess}(f)$ with $max(\lambda_k, \epsilon)$, without changing the corresponding eigenvectors. In this case the results demonstrate a clear advantage of second-order optimizers with respect to gradient descent.

\end{document}